\documentclass[a4paper,11pt]{article}
\usepackage{pos}

\title{A string-inspired running-vacuum-model of cosmology, primordial-black-hole dark matter, and the current tensions in cosmological data}
\ShortTitle{Stringy RVM and Cosmological Tensions}

\author*[a, b]{Nick E. Mavromatos}
\affiliation[a]{National Technical University of Athens, School of Applied Mathematical and Physical Sciences, Department of Physics, 9 Iroon Polytechniou Street, Zografou Campus GR157 80, Athens, Greece}

\affiliation[b]{King's College London, Department of Physics, Strand, London WC2R 2LS, UK}

\emailAdd{mavroman@mail.ntua.gr}

\abstract{I review a cosmological model based on string-inspired gravity with anomalies and torsion, which leads to a cosmology of running-vacuum-model (RVM) type. It is argued that such a model can lead to observable, in principle, deviations  from the standard $\Lambda$CDM paradigm of cosmology at late (modern) eras of the Universe evolution, 
contributing to an alleviation of the currently observed cosmological $H_0$ and structure-growth tensions, but also to a potential enhancement of the density of primordial black holes produced during inflation. The latter can thus play a r\^ole as dark matter components, and leave observable imprints in gravitational-wave patterns during the radiation-dominance era of the Universe.}

\FullConference{%
  Corfu Summer Institute 2022 "School and Workshops on Elementary Particle Physics and Gravity",\\
  28 August - 1 October, 2022\\
  Corfu, Greece}

\begin{document}
\maketitle

\section{Introduction}\label{sec:intro}

Although the standard concordance model of Cosmology, aka $\Lambda$CDM, which includes a Cold Dark Matter (CDM) component and a positive cosmological constant $\Lambda > 0$ era dominating the current epoch of the Universe evolution, and thus contributing to a late Universe acceleration, fits the plethora of the cosmological data very well~\cite{Planck}, nonetheless there are recent tensions in the data. Those are associated with a discrepancy between the directly measured value of the current-era Hubble parameter, $H_0$, using nearby galaxy data (Cepheids etc.)~\cite{tensions}, and the prediction for $H_0$ coming from $\Lambda$CDM-based fits of data using Cosmic Microwave Background (CMB) data of the Planck satellite (including also analysis of Baryon acoustic oscillation (BAO) data and gravitational lensing fits)~\cite{Planck}. Similar tensions appear in galaxy growth data, e.g. in the so-called $\sigma_8$ or  $\sigma_{12}$ parameters, that is the variances $\sigma(R)$ of linear matter fluctuations within spheres of radii $R=8h^{-1}$ Mpc and $R=12$ Mpc, respectively (where $h$ is the reduced Hubble parameter). Although it is still not clear whether such tensions can be explained by mundane astrophysical explanations, or are due to the current statistical uncertainties, thus, subject to potential  disappearance in future, more accurate, measurements~\cite{freedman}, nonetheless they triggered, due to their persisting nature, an enormous theoretical interest, as they might imply true deviations from the $\Lambda$CDM paradigm~\cite{models}, and hence new fundamental physics. 

One of the cosmological frameworks that claims to offer potential alleviations of both types of tensions, the $H_0$, as well as the $\sigma_8$, is the so-called running-vacuum-model (RVM) of cosmology~\cite{rvm,Solarvm}. According to this framework, the entire evolution of the Universe can be explained in terms of a smooth running~\cite{lima} with the cosmic time $t$ of the cosmological vacuum energy $\rho_{\rm RVM}(t)$. Due to {\it general covariance}, this quantity admits a perturbative expansion on {\it even} powers of the Hubble parameter $H$ :\footnote{Such an RVM evolution can also explain the thermodynamical aspects of the Universe, including an explanation of its large entropy production~\cite{rvmthermal}.}
\begin{align}\label{rvmener}
\rho_{\rm RVM} (H) = \frac{\Lambda (H^2)}{\kappa^2} = \frac{3}{\kappa^2} \Big( c_0 + \nu\, H^2 + \alpha\, \frac{H^4}{H_I^2} + \dots \Big), 
\end{align}  
where $\kappa^2 = 8\pi \rm G = M_{\rm Pl}^{-2}$ is the four-dimensional gravitational constant (with G the Newton's constant, and $M_{\rm Pl}=2.4 \times 10^{18}~{\rm GeV}$ the reduced Planck mass), $c_0, \nu, \alpha$ are real constant (and positive in the conventional RVM framework) dimensionless coefficients (with $H_I \sim 10^{-5} \, M_{\rm Pl}$ the inflationary scale, e.g. as measured by Planck Coll.~\cite{Planck}), and the $\dots$ denote higher even powers of $H$. The constant $c_0 \kappa^{-4}$ is set to the current value of the observed cosmological constant~\cite{Planck}, $c_0 \, \kappa^{-4} \sim 10^{-122} M_{\rm Pl}^4$.\footnote{In the initial formulation of the RVM~\cite{rvm}, in which the cosmic evolution is viewed as a ``renormalization-group (RG)''-type evolution, with the RG scale being given by $H^2$, such a constant arises as an integration constant when one computes the vacuum energy $\rho_{\rm RVM}$ from the postulated ``RG-beta-like-function'' formulation of the RVM evolution:
$$ \frac{d\rho_{{\rm RVM}}}{d{\rm ln}(H^2)}  = 
\frac{1}{4\pi^2} \, \sum_i \Big(a_i M_i^2 H^2 + b_i H^4 + c_i \frac{H^6}{M_i^2} \dots \Big)\,,$$ 
where the dimensionless coefficients $a_i,b_i$ are linked to loop contributions of quantum matter and radiation fields (fermions or bosons) of masses $M_i^2$. In general, on the right hand-side of the above relation, there are also contributions proportional to $\dot H$ (where the dot denotes derivative with respect to the cosmic time $t$). However, such contributions can be expressed 
in terms of the cosmic deceleration parameter $q=- \frac{\ddot{a}\, a }{(\dot a)^2} $ as $\dot{H}=-(q+1) H^2$. Making the physically relevant approximation that 
the deceleration parameter takes on  approximately constant values during the various epochs of the Universe evolution, $q \simeq 1,1/2,-1$, as we move from radiation- into the matter- and eventually dark-energy -dominated epochs, respectively, we can see that the modifications introduced by $\dot{H}$ are expressed in terms of $H^2$, thereby justifying the assumption \eqref{rvmener} that the RVM energy density is, to a good approximation, a function of even powers of $H$.}

It should be remarked that the entire history of the Universe can be described by the truncation of the expansion \eqref{rvmener} to $H^4$ terms~\cite{lima}, although it should be noted that quantum field theory computations in RVM spacetimes lead to $H^2$ and $H^6$ terms, but not $H^4$~\cite{qftrvm}. The latter can be produced by condensates of primordial gravitational tensor modes (of chiral gravitational-wave (GW) type) in string-inspired RVM models (stringy RVM), characterised by torsion and Chern-Simons gravitational anomalies~\cite{jackiw}, as we discussed in \cite{bms,ms} and review in this talk. It should be mentioned, though, that the cosmological evolution of the stringy RVM are characterised by different phases, in which the coefficient $\nu$ is negative during the inflationary era, and positive afterwards. Moreover, in the stringy RVM $c_0 =0$, since a cosmological constant is incompatible with either  perturbative (scattering S-matrix)~\cite{smatrix} or non-perturbative string theory (swampland criteria)~\cite{swamp}. In our stringy RVM, approximate cosomological constant terms may arise either during the early RVM inflationary phase or at late eras, as a result of the formation of appropriate condensates~\cite{bms,ms}, which eventually disappear, and thus any de Sitter epoch appears as {\it metastable}. Such a metastability feature of the de Sitter phase seems to be also corroborated by local quantum field theory computations and vacuum renormalization within a conventional RVM expanding universe background framework~\cite{qftrvm}.  

The RVM vacuum fluid is characterised by a de-Sitter type equation of state~\cite{rvm,Solarvm,lima}:
\begin{align}\label{rvmeos}
p_{\rm RVM} = - \rho_{\rm RVM}\,.
\end{align}
It should be noted that \eqref{rvmeos} is essentially valid at early Universe epochs, when the vacuum gravitational degrees of freedom are dominant. As discussed in \cite{ms}, and will be reviewed explicitly below, this is the case of the stringy RVM at early epochs. In radiation and matter-dominance eras, one should consider the contributions to the energy density of such excitations, $\rho_m$ ($m$=radiation, matter), on top of the above vacuum contributions, which thus lead to a total energy density of the form $\rho_{\rm total} = \rho_{\rm RVM} + \rho_m $. The equations of state of such total contributions coincide approximately with the equation of state of the respective dominant component in the corresponding era. Such a result has been confirmed by detailed quantum field theory computations in such RVM spacetime backgrounds~\cite{qfteos}, including appropriate renormalization  of the vacuum energy~\cite{qftrvm}. 

At late eras, the RVM framework is argued~\cite{rvmpheno} (see also \cite{tsiapi}) to provide a phenomenology consistent with the plethora of the cosmological data available today~\cite{Planck}, including cosmic structure formation~\cite{rvmstruct}. However, the RVM framework also indicate observable deviations from the $\Lambda$CDM paradigm, due to the $\nu H^2$ term, which dominates the late epochs of the cosmological evolution. Fitting the current-era data with the RVM evolution framework leads to the estimate  $\nu = \mathcal O(10^{-3})$. Curiously enough, the same estimate for $\nu$ is obtained by requiring consistency of the RVM framework with the Big-Bang-Nucleosynthesis (BBN) data~\cite{bbnrvm}, upon assuming that the observed~\cite{Planck} current-era dark energy is of RVM type. 
On the other hand, the $H^4$ (and higher order) terms in \eqref{rvmener}, which are dominant during the early eras of the cosmic evolution, drive an RVM inflation without the need for fundamental inflaton fields~\cite{lima}. 

Moreover, there are claims~\cite{rvmtensions} that a modified RVM, called RVM type II, allowing for a mild phenomenological dependence of the gravitational constant on the cosmic time $t$,  $\kappa \to \kappa (t)$, can simultaneously alleviate the 
$H_0$ and $\sigma_8$ tensions observed in late-cosmology data~\cite{tensions}. As we shall argue below, the stringy RVM also provides a resolution of such tensions, in similar spirit to the RVM type II. There is an extra mild dependence on cosmic time in the stringy RVM, as compared to the conventional RVM framework, as a consequence of quantum-gravity- induced logarithmic corrections $H^2{\rm ln}(H^2)$ to the modern era vacuum energy (in fact such corrections should characterise any gravitational theory, being due to quantum one-loop contributions to the effective action)~\cite{mavrophil,ms}. Under some circumstances, which we shall discuss, such corrections can dominate the corresponding ones induced by appropriately renormalised quantum matter fields in the expanding-universe spacetime background~\cite{qftrvm}.

The structure of the talk is the following: in the next section \ref{sec:stringyRVM}, we formulate the stringy RVM, as a gravitational model, inspired from string theory, which is characterised by torsion and Chern-Simons type gravitational anomalies at early epochs. In section \ref{sec:rvminfl}, we discuss how, upon condensation of primordial chiral GW modes,
one obtains inflation of RVM type, without the need for external inflaton fields, but only as a result of non-linearities of the RVM vacuum. In section \ref{sec:pbh}, we discuss briefly how in such string inspired models one may have an enhanced production of primordial black holes (pBH), which in this way could play a r\^ole of a Dark Matter (DM) component, also affecting the profile of the GW produced during the radiation era after exit from RVM inflation. Finally, in section \ref{sec:now} we discuss how in the current era, the stringy RVM could contribute to a possible alleviation of the current-era $H_0$ tension, as well as tensions associated with galactic growth parameters, in particular the $\sigma_{12}$ in this case~\cite{gms}. Conclusions and outlook are given in section \ref{sec:concl}. 

\section{String-inspired  Chern-Simons cosmological model of RVM type: Stringy RVM}\label{sec:stringyRVM}

The string-inspired cosmology model of \cite{bms,ms} is based on the bosonic part of the massless gravitational multiplet of the (closed sector of the) underlying microscopic superstring theory, which constitutes also the ground state~\cite{string}. This part consists of the fields of graviton (spin-2 symmetric tensor), dilaton (scalar, spin-0 field) and  spin-1 antisymmetric tensor (or Kalb-Ramond (KR)) gauge field, $B_{\mu\nu}=-B_{\nu\mu}$. That the latter is a gauge field follows from the invariance of the 
vertex $\sigma$-model operator corresponding to the $B_{\mu\nu}$ field excitation:
\begin{align}\label{bvertex}
V_B = \int_{\Sigma^{(2)}} \, d^2 \xi \, \epsilon^{AB} \, B_{\mu\nu}(X) \, \partial_A X^\mu \,  \partial_B X^\nu\,,
\end{align}
under the following Abelian (U(1)) gauge transformation in the target space of the string,\footnote{The invariance of 
\eqref{bvertex} under the transformation \eqref{kru1} follows from the application of Stokes theorem on the integral over the boundaryless world sheet  $\Sigma^{(2)}$.} 
 
\begin{align}\label{kru1}
B_{\mu\nu}(X) \, \Rightarrow \, B_{\mu\nu}(X) + \partial_\mu \theta_\nu(X) - \partial_\nu \theta_\mu (X)\,, 
\end{align}
where $A, B = 1,2$ are indices on the two-dimensional world-sheet manifold, $\Sigma^{(2)}$, with the topology of a sphere (without a boundary), with coordinates $\xi^1=\tau,\, \xi^2= \sigma$, and  
$\epsilon^{AB}=-\epsilon^{BA} $ is the world-sheet covariant Levi-Civita antisymmetric symbol. The world-sheet zero modes of the two-dimensional scalar ields $X^\mu (\xi)$ correspond to the target spacetime coordinates. After compactification to four dimensions, which we restrict our attention to for our purposes here, the Greek indices $\mu,\nu=0, \dots 3$ denote (3+1)-dimensional target spacetime indices. 

As a consequence of the gauge symmetry \eqref{kru1}, the target-spacetime effective action deriving from the string theory at hand depends only on the field strength of the KR field $B_{\mu\nu}$:
\begin{align}\label{krH}
 \mathcal{H}_{\mu\nu\rho} = \kappa^{-1}\, \partial_{[\mu} B_{\nu\rho]}\,, \quad \mu, \nu, \rho =0, \dots 3\,,
\end{align}  
where $[\dots ]$ denotes full antisymmetry of the corresponding indices. The Green-Schwarz anomaly cancellation mechanism~\cite{gs, string} implies a modification of \eqref{krH} via the Lorentz (L) and Yang-Mills gauge (Y) Chern-Simons three forms:
\begin{align}\label{csterms}
\mathbf{{\mathcal H}} &= \kappa^{-1}\, \mathbf{d} \mathbf{B} + \frac{\alpha^\prime}{8\, \kappa} \, \Big(\Omega_{\rm 3L} - \Omega_{\rm 3Y}\Big),  \nonumber \\
\Omega_{\rm 3L} &= \omega^a_{\,\,c} \wedge \mathbf{d} \omega^c_{\,\,a}
+ \frac{2}{3}  \omega^a_{\,\,c} \wedge  \omega^c_{\,\,d} \wedge \omega^d_{\,\,a},
\quad \Omega_{\rm 3Y} = \mathbf{A} \wedge  \mathbf{d} \mathbf{A} + \mathbf{A} \wedge \mathbf{A} \wedge \mathbf{A},
\end{align}
where $\alpha^\prime = M_s^{-2}$ is the Regge slope of the string, with $M_s$ the string-mass scale, which is a free parameter in string theory, implying that in general $\sqrt{\alpha^\prime} \ne \kappa$. In \eqref{csterms}
we used differential form language~\cite{Eguchi} for notational brevity, with $\wedge$ denoting the usual exterior (``wedge'') product among differential forms, such that  ${\mathbf f}^{(k)} \wedge {\mathbf g}^{(\ell)} = (-1)^{k\, \ell}\, {\mathbf g}^{(\ell)} \wedge {\mathbf f}^{(k)}$, where ${\mathbf f}^{(k)}$, and ${\mathbf g}^{(\ell)}$ are $k-$ and $\ell-$ forms, respectively. The quantity $\mathbf{A}$ in \eqref{csterms} is the Yang-Mills potential (gauge field) one form, while $\omega^a_{\,\,b}$ is the spin connection one form (the Latin indices $a,b,c,d$ are tangent space (i.e. Lorentz group SO(1,3)) indices). 

The low-energy gravitational theory corresponding to the bosonic massless gravitational multiplet of the superstring is given by~\cite{string}:\footnote{Our conventions and definitions used throughout this work are those of \cite{bms}, that is: signature of metric $(+, -,-,- )$, Riemann Curvature tensor
$R^\lambda_{\,\,\,\,\mu \nu \sigma} = \partial_\nu \, \Gamma^\lambda_{\,\,\mu\sigma} + \Gamma^\rho_{\,\, \mu\sigma} \, \Gamma^\lambda_{\,\, \rho\nu} - (\nu \leftrightarrow \sigma)$, Ricci tensor $R_{\mu\nu} = R^\lambda_{\,\,\,\,\mu \lambda \nu}$, and Ricci scalar $R = R_{\mu\nu}g^{\mu\nu}$.}

\begin{align}\label{sea}
S_B  =\; \int d^{4}x\sqrt{-g}\Big( \dfrac{1}{2\kappa^{2}} [-R + 2\, \partial_{\mu}\Phi\, \partial^{\mu}\Phi] - \frac{1}{6}\, e^{-4\Phi}\, {\mathcal H}_{\lambda\mu\nu}{\mathcal H}^{\lambda\mu\nu}  + \dots \Big),
\end{align}
where the $\dots$ denote higher-derivative terms and perhaps string loop-induced dilaton potentials. As already mentioned, we do not include a bare cosmological constant in the action \eqref{sea}, as this would be incompatible with both, perturbative~\cite{smatrix} and non-perturbative string theory~\cite{swamp}. Moreover, in the model of \cite{bms,ms}, as a self-consistent solution of the equations of motion~\cite{bms} (upon including minimisation of appropriate dilaton potentials, with their minima corresponding to zero cosmological constant), the dilaton $\Phi$ is taken to be a constant, 
 which can be set to zero, thereby fixing the normalization of the string coupling $g=g_s e^\Phi\Big|_{\Phi=0}\, \Rightarrow \, g=g_s$. This is what we assume in our analysis below.

The modified definition (\ref{csterms}) of the field strength $\mathcal H$
leads to the Bianchi identity~\cite{string}
\begin{equation}\label{modbianchi}
\mathbf{d} \mathbf{{\mathcal H}} = \frac{\alpha^\prime}{8 \, \kappa} {\rm Tr} \Big(\mathbf{R} \wedge \mathbf{R} - \mathbf{F} \wedge \mathbf{F}\Big)
\end{equation}
with $\mathbf{F} = \mathbf{d} \mathbf{A} + \mathbf{A} \wedge  \mathbf{A}$ the Yang-Mills field strength two form,  and $\mathbf{R}^a_{\,\,b} = \mathbf{d} \omega^a_{\,\,b} + \omega^a_{\,\,c} \wedge \omega^c_{\,\,b}$ the curvature two form. The trace (Tr) is taken over both gauge and Lorentz group indices.
The non zero quantity on the right hand side  of \eqref{modbianchi} is the ``mixed (gauge and gravitational) quantum anomaly''~\cite{alvwitt}. In the usual tensor notation,  
the Bianchi identity (\ref{modbianchi}) is expressed as:
\begin{equation}\label{modbianchi2}
 \varepsilon_{abc}^{\;\;\;\;\;\mu}\, {\mathcal H}^{abc}_{\;\;\;\;\;\; ;\mu}
 -  \frac{\alpha^\prime}{32\, \kappa} \, \sqrt{-g}\, \Big(R_{\mu\nu\rho\sigma}\, \widetilde R^{\mu\nu\rho\sigma} -
\mathbf F_{\mu\nu}\, \widetilde{\mathbf F}^{\mu\nu}\Big) \equiv \sqrt{-g}\, {\mathcal G}(\omega, \mathbf{A}) =0,
\end{equation}
where the semicolon denotes covariant derivative with respect to the standard
Christoffel connection, and
\begin{equation}\label{leviC}
\varepsilon_{\mu\nu\rho\sigma} = \sqrt{-g}\,  \epsilon_{\mu\nu\rho\sigma}, \quad \varepsilon^{\mu\nu\rho\sigma} =\frac{{\rm sgn}(g)}{\sqrt{-g}}\,  \epsilon^{\mu\nu\rho\sigma},
\end{equation}
with $\epsilon^{0123} = +1$, {\emph etc.}, are the gravitationally covariant Levi-Civita tensor densities, totally antisymmetric in their indices.
The symbol
$\widetilde{(\dots)}$
over the curvature or gauge field strength tensors denotes the corresponding duals, defined as
\begin{align}\label{duals}
\widetilde R_{\mu\nu\rho\sigma} = \frac{1}{2} \varepsilon_{\mu\nu\lambda\pi} R_{\,\,\,\,\,\,\,\rho\sigma}^{\lambda\pi}, \quad \widetilde{\mathbf F}_{\mu\nu} = \frac{1}{2} \varepsilon_{\mu\nu\rho\sigma}\, \mathbf F^{\rho\sigma}.
\end{align}

The anomaly ${\mathcal G}(\omega, \mathbf{A})$ is an exact one loop result~\cite{alvwitt}, which implies that the Bianchi identity (\ref{modbianchi2}) is a constraint of the quantum theory. Such a constraint is implemented as a $\delta$-functional constraint in the quantum path integral of the action (\ref{sea}) (upon setting $\Phi=0$) over the fields ${\mathcal H}$, $\mathbf{A}$, and $g_{\mu\nu}$. 
The $\delta$ functional constraint is then expressed in terms of a (canonically normalised) Lagrange multiplier (pseudoscalar) field~\cite{kaloper,svrcek} $b(x)$ :
\begin{align}\label{delta}
&\Pi_{x}\, \delta\Big(\varepsilon^{\mu\nu\rho\sigma} \, {{\mathcal H}_{\nu\rho\sigma}(x)}_{; \mu} - {\mathcal G}(\omega, \mathbf{A}) \Big)
\Rightarrow  \nonumber \\ &\int {\mathcal D}b \, \exp\Big[i \, \,\int d^4x \sqrt{-g}\, \frac{1}{\sqrt{3}}\, b(x) \Big(\varepsilon^{\mu\nu\rho\sigma }\, {{\mathcal H}_{\nu\rho\sigma}(x)}_{; \mu} - {\mathcal G}(\omega, \mathbf{A}) \Big) \Big] \nonumber \\
&= \int {\mathcal D}b \, \exp\Big[-i \,\int d^4x \sqrt{-g}\, \Big( \partial ^\mu b(x) \, \frac{1}{\sqrt{3}} \, \epsilon_{\mu\nu\rho\sigma} \,{\mathcal H}^{\nu\rho\sigma}  + \frac{b(x)}{\sqrt{3}}\, {\mathcal G}(\omega, \mathbf{A}) \Big)\Big]
\end{align}
where in the second equality we have performed partial integration, assuming that the KR field strength dies out at spatial infinity. Inserting (\ref{delta})
into the path integral of the action (\ref{sea}), and integrating over the ${\mathcal H}$ field,
which for this action (which is truncated to second order in a target-space derivative expansion of the string effective action~\cite{string}) can be done exactly, as $\mathcal{H}$ is a non-propagating field,  
one obtains a dynamical pseudoscalar (axion-like) field, $b(x)$, coupled to the Chern-Simons (CS) mixed anomaly ${\mathcal G}(\omega, \mathbf{A})$~\cite{kaloper,svrcek}:
\begin{align}\label{sea3}
S^{\rm eff}_B =&\; \int d^{4}x\sqrt{-g}\Big[ -\dfrac{1}{2\kappa^{2}}\, R + \frac{1}{2}\, \partial_\mu b \, \partial^\mu b +  \sqrt{\frac{2}{3}} \, \frac{\alpha^\prime}{96\, \kappa} \, b(x) \, \Big(R_{\mu\nu\rho\sigma}\, \widetilde R^{\mu\nu\rho\sigma} - \mathbf{F}_{\mu\nu}\, \widetilde{\mathbf F}^{\mu\nu}\Big) + \dots \Big],
\end{align} 
where the dots $\dots$ denote gauge kinetic, as well as higher-derivative, terms appearing in the string effective action, that we ignore for our discussion here. The field $b(x)$ is called gravitational, or KR, or, in modern string-theory language, string-model independent axion~\cite{svrcek}, as it characterises all string theories, independent of their types of compactified geometries. In string theory, there are of course many more axions, arising from compactification, which may lead to a rich phenomenology~\cite{arvanitaki} and cosmology~\cite{marsh}. We shall discuss a potential r\^ole of the compactification axions in our context later on. 

We thus observe that the KR axion field couples to the gravitational and gauge fields via the mixed anomaly. The corresponding interaction coupling is the inverse of the KR-axion coupling parameter $f_b$ (of mass dimension +1), which in view of \eqref{sea3}, is given by 
\begin{align}\label{fbdef}
f_b^{-1} \equiv  \sqrt{\frac{2}{3}} \, \frac{\alpha^\prime}{96\, \kappa} \,.
\end{align}
This interaction is Charge-conjugation-symmetry (C) conserving but Parity (P)  and Time-reversal (T) violating, and hence in view of the overall CPT invariance of the quantum theory, also CP violating. In fact, the mixed-anomaly  term is a total derivative
\begin{align}\label{pontryaginA}
&\sqrt{-g} \, \Big(R_{\mu\nu\rho\sigma}\, \widetilde R^{\mu\nu\rho\sigma} - \mathbf F_{\mu\nu}\, \widetilde{\mathbf F}^{\mu\nu} \Big) = \sqrt{-g} \, {\mathcal K}_{\rm mixed}^\mu (\omega, \mathbf A)_{;\mu} = \partial_\mu \Big(\sqrt{-g} \, {\mathcal K}_{\rm mixed}^\mu (\omega, \mathbf{A}) \Big) \nonumber \\
&= 2 \, \partial_\mu \Big[\epsilon^{\mu\nu\alpha\beta}\, \omega_\nu^{ab}\, \Big(\partial_\alpha \, \omega_{\beta ab} + \frac{2}{3}\, \omega_{\alpha a}^{\,\,\,\,\,\,\,c}\, \omega_{\beta cb}\Big) - 2 \epsilon^{\mu\nu \alpha\beta}\, \Big(A^i_\nu\, \partial_\alpha A_\beta^i + \frac{2}{3} \, f^{ijk} \, A_\nu^i\, A_\alpha^j \, A_\beta^k \Big)\Big],
\end{align}
with Latin letters $i,j,k$ being gauge group indices, and $\sqrt{-g}\, {\mathcal K}_{\rm mixed}^\mu (\omega, \mathbf A)$ denoting the mixed (gauge and gravitational) anomaly current (or topological) density.
The model \eqref{sea3} constitutes a CS modification of general relativity (GR)~\cite{jackiw}.\footnote{This is a quartic-order derivative modification of GR, which however does not lead to graviton ghosts. Another invariant, which also does not lead to graviton ghosts in (3+1)-dimensions are the so-called Gauss-Bonnet quadratic-curvature terms. 
In view of setting $\Phi=0$ in our case, such terms are absent, given the total derivative nature of the GB terms. As discussed in \cite{bms}, the CS modifications lead to negative contributions to the stress tensor, in similar spirit to the GB terms, a feature that is relevant for the existence of  dilaton scalar hair in the respective black-hole solutions of GB gravity~\cite{kanti}. Notably, the local solutions of the action \eqref{sea3} contain rotating black holes~\cite{yunes,dorlis}, with back-reacting pseudoscalar hair, whose existence is also due to the violation of appropriate energy conditions~\cite{dorlis}.}

The presence of axion-like degrees of freedom in the spectrum of this string-inspired gravitational theory is in agreement~\cite{kaloper} with the r\^ole of the field strength $\mathcal H_{\mu\nu\rho}$ as torsion~\cite{torsion}, a result that follows by absorbing the quadratic $\mathcal H^2$ terms in \eqref{sea3} into a generalised curvature scalar term $\overline{R}(\overline \Gamma)$ defined with respect to a contorted generalised comnnection $\overline \Gamma^\rho_{\,\,\mu\nu} \ne \Gamma^\rho_{\,\,\nu\rho}$:
\begin{align}\label{torcon}
{\overline \Gamma}_{\mu\nu}^{\rho} = \Gamma_{\mu\nu}^\rho + \frac{\kappa}{\sqrt{3}}\, {\mathcal H}_{\mu\nu}^\rho  \ne {\overline \Gamma}_{\nu\mu}^{\rho}~,
\end{align}
with $\mathcal H^\rho_{\,\,\mu\nu} = - \mathcal H^\rho_{\,\,\nu\mu}$ playing the r\^ole of the contorsion~\cite{torsion}, and 
$\Gamma_{\mu\nu}^\rho = \Gamma_{\nu\mu}^\rho$ the torsion-free Christoffel symbol.\footnote{Exploiting local field redefinition ambiguities~\cite{string,kaloper}, which do not affect the perturbative scattering amplitudes, and thus the S-matrix, one may extend the above conclusion to ${\mathcal O}(\alpha^\prime)$ effective low-energy action, which includes terms of quartic-order in derivatives.} The three form $\mathcal H_{\mu\nu\rho}$ corresponds to a totally antisymmetric torsion in a purely bosonic gravity model. Par contrast, in the Einstein-Cartan contorted field theories, like the contorted quantum electrodynamics ({\it torsQED}) discussed in~\cite{kaloper}, the torsion has fermionic origin. Nonetheless, the origin of the axion-like degrees of freedom associated with the torsion in both classes of theories is similar. In {\it torsQED}, 
it is the conservation of the torsion charge that needs to be guaranteed order by order in perturbation theory that introduces
a Bianchi-like constraint into the path-integrated quantum theory~\cite{kaloper}:
\begin{align}\label{ECtors}
\mathbf d ^\star \mathbf  S =0\,, 
\end{align}
where $^\star$ denotes the dual Hodge operation in differential geometry~\cite{Eguchi}. The one form 
$\mathbf S = ^\star \mathbf T$, is the Hodge dual of the torsion three form $\mathbf T$ (in components $\mathbf T_{\mu\nu\rho}$) and is thus  associated with 
 the totally antisymmetric component of the torsion. Classically $\mathbf S$ is proportional to the axial fermion current, 
 $\mathbf S \propto \mathbf J^5$, where $J^{5\mu } = \overline \psi \, \gamma^\mu \, \gamma^5 \, \psi$, with $\psi (x)$ denoting generic fermion fields. The constraint \eqref{ECtors} is implemented in the relevant path integral by means of an axion-like Lagrange multiplier field which plays the same r\^ole as the KR (or string-model independent) axion, which is the Lagrange multiplier implementing the Bianchi constraint \eqref{modbianchi2} (or, equivalently, \eqref{modbianchi}).

At this point an important remark is in order. Notice that the modifications (\ref{csterms}) and the right-hand-side of the Bianchi (\ref{modbianchi}) contain the {\it torsion-free} spin connection. In fact, it can be shown~\cite{hull} that any potential contributions from the torsion $\mathcal{H}_{\mu\nu\rho}$ three form to the anomaly equation can be removed by adding to the string effective action appropriate counterterms order by order in perturbation theory.

\section{Stringy RVM inflation from primordial gravitational waves}\label{sec:rvminfl}

In the cosmology model of \cite{bms,ms} it is assumed that in the early Universe only fields from the bosonic string gravitational multiplet appear as external fields in the effective action. That is, the dynamics of the primordial Universe is given by
\eqref{sea3}, assuming $\Phi=0$ and no gauge ({\it i.e.} $\mathbf A=0$) or other matter fields.\footnote{We note that such cosmologies based on the bosonic massless gravitational multiplets of strings have also been considered in \cite{aben}, in the context of the so-called non-critical string cosmologies. In our case, the effective model is assumed to stem from an underlying  critical string theory, although the model can be adapted to include non-critical strings. In the latter case, the dilaton cannot be set to a constant, and appropriate dilaton potentials have to be explicitly considered.} 
As we shall discuss below, all chiral fermion and other matter or radiation fields are produced at the end of the inflationary era in this model, which is of RVM type, not requiring external inflatons~\cite{lima}. 
In fact, in \cite{ms} it was also assumed that in this model, immediately after the Big-Bang, one may have  a local supersymmetry (supergravity) phase, where gravitinos, $\psi_\mu$, the spin 3/2 supersymmetric partners of gravitons, are also present, and they condense to break the supergravity dynamically via a double-well potential of the gravitino condensate 
$\sigma=\langle \overline{\psi}_\mu \, \psi^\mu \rangle$~\cite{houston}. In the broken supergravity phase, the gravitino and its condensate acquire large masses (close to Planck) and they decouple from the spectrum. In fact in this scenario, there could be a primordial first hill-top inflation~\cite{ellisinfl}, which is not necessarily slow roll, as it does not lead to observable consequences other than  homogeneity and isotropy, which can then be used to describe quantitatively the transition, via tunnelling~\cite{ms}, of the isotropic and homogeneous system of massless gravitons and KR axions to the RVM second inflationary phase~\cite{ms}, that we now proceed to discuss.

After the dynamical local supersymmetry breaking, and the exit from the first hill-top inflation, gravitational waves can form, as a consequence of a lifting of the degeneracy of the vacua of the double-well potential of the gravitino condensate due to 
dynamical percolation effects in the early Universe, which result in asymmetric occupation numbers of the two vacua~\cite{ross}. Such a lifting may lead to the formation of domain walls, whose asymmetric collapse and/or collisions can lead to the formation of primordial {\it chiral} (left-right asymmetric)  gravitational waves (GW). In the scenario of \cite{ms}, such GW can condense leading in turn to non-trivial condensates of the CS gravitational anomaly terms. 

If we assume $\mathcal N(t)$ sources of such chiral primordial GW~\cite{mavlorentz}, then it can be shown, following~\cite{lyth}, that the GW induced CS condensate can be estimated as (assuming initially an approximately constant $H$, {\it i.e.} inflation, which we shall argue later that it is a consistent solution of the pertinent cosmic evolution):
 \begin{align}\label{condensateN2}
\langle R_{\mu\nu\rho\sigma} \, \widetilde R^{\mu\nu\rho\sigma} \rangle 
=\frac{\mathcal N(t)}{\sqrt{-g}}  \, \frac{1.1}{\pi^2} \, 
\Big(\frac{H}{M_{\rm Pl}}\Big)^3 \, \mu^4\, \frac{\dot b(t)}{M_s^{2}} \equiv n_\star \, \frac{1.1}{\pi^2} \, 
\Big(\frac{H}{M_{\rm Pl}}\Big)^3 \, \mu^4\, \frac{\dot b(t)}{M_s^{2}}~,
\end{align} 
where the overdot denotes derivative with respect to the cosmic time, and $n_\star \equiv \mathcal N(t)/\sqrt{-g}$ is the proper density of sources of GW, which we may assume to be approximately constant during the RVM inflation for simplicity and concreteness. The quantity $\mu$ is the UV cutoff of the effective low-energy theory, which serves as an upper bound for the momenta of the graviton modes that are integrated over in the computation of the condensate \eqref{condensateN2} in  the presence of chiral GW. To arrive at \eqref{condensateN2}, we have assumed isotropy and homogeneity of the string Universe, which can be guaranteed by the first hill-top inflation in the pre-RVM-inflationary scenario of \cite{ms}, as discussed in the beginning of this section.\footnote{It should be stressed that the 
estimate \eqref{condensateN2} is a valid estimate only in the field theory low-energy limit of the corresponding string theory. Unfortunately, as the estimate relies on the dominant UV physics near the cutoff $\mu$, in the case of microscopic string theory models it is the entire towers of massive string states that contribute, which makes an accurate estimation of the CS anomaly condensate not possible at present, apart from ensuring its non vanishing value.}

From the anomaly equation \eqref{pontryaginA}, then, on imposing homogeneity and isotropy, we obtain
\begin{align}\label{anomevol}
\frac{d}{dt}  <\mathcal K^0>  & + 3 \, H\, <\mathcal K^0>  \,
\simeq \, n_\star \,  \frac{1.1}{\pi^2} \, 
\Big(\frac{H}{M_{\rm Pl}}\Big)^3 \, \mu^4\, \frac{\dot b(t)}{M_s^2} \, .
\end{align}
Using the KR-axion equation of motion obtained from \eqref{sea3}
we then have~\cite{bms}
\begin{align}\label{eqmotb}
\dot b (t) =\sqrt{\frac{2}{3}} \frac{\alpha^\prime}{96\, \kappa}\, <\mathcal K^0>\,,
  \end{align}
 On making 
 the reasonable assumption that the UV cutoff $\mu$ of the string-effective low-energy field theory is identified with the string Mass scale~\cite{mavlorentz}, 
 {\it i.e.}
 \begin{align}\label{cutoff}
 \alpha^\prime = \mu^{-2}\,, 
 \end{align}
 we obtain from \eqref{anomevol} that a consistent solution for the temporal component of the anomaly current implies
 \begin{align}\label{anomevol2}
 <\mathcal K^0>  & \simeq {\rm constant} \ne 0\,.
 \end{align}
 Then,  $\langle \mathcal K^0 \rangle \ne 0$ drops out of the Eq.~\eqref{anomevol}, 
 and we obtain~\cite{mavlorentz}
 \begin{align}\label{valuenstar}
n_\star^{1/4} \, \sim \,  7.6 \times \Big(\frac{M_{\rm Pl}}{H}\Big)^{1/2}\,.
 \end{align}
 If one uses the Planck cosmological data~\cite{Planck} to fix $H \simeq H_I $ during inflation to the range of values 
$H_I / M_{\rm Pl} \lesssim 10^{-5}$, then we obtain~\cite{mavlorentz} 
\begin{align}\label{nstarval}
n_\star \gtrsim 3.3\times 10^{13}\,, 
\end{align} 
which gives the range of values of the proper number density  of macroscopic sources of primordial chiral GW that are 
 needed to produce an approximately constant gravitational anomaly condensate in the context of the CS cosmology of \cite{bms}. 
 
The explicit computations of \cite{bms,ms,mavlorentz} have shown that the total vacuum energy density during inflation assumes an RVM form \eqref{rvmener}:
\begin{align}\label{totalenerden}
\rho^{\rm total}_{\rm vac} =  -\frac{1}{2}\, \epsilon \, M_{\rm Pl}^2\, H^2 + 4.3 \times 10^{10} \, \sqrt{\epsilon}\, 
\frac{|\overline b(0)|}{M_{\rm Pl}} \, H^4\,,
\end{align}
where we have used the following parametrisation of the approximately constant $\dot b$ axion background (\eqref{eqmotb}, \eqref{anomevol2}) during inflation: 
\begin{align}\label{axionbackgr}
b(t)=\overline{b}(t_0) + \sqrt{2\epsilon} \, H \, (t-t_0) \, M_{\rm Pl}\,,  \quad \overline{b}(t_0) < 0\,,
\end{align}
where $0 < \epsilon < 1 $ is a phenomenological parameter, and the time $t_0$ corresponds to the onset of the RVM inflation, implying that $\overline b(t_0)$ plays the r\^ole of a boundary condition for the KR axion.  In arriving at the estimates of \eqref{totalenerden} we took into account the fact that the assumption on the approximate de Sitter (positive-cosmological-constant) nature of the condensate, so as to lead to an approximately constant $H \simeq H_I$ during the RVM inflation, requires~\cite{bms,ms}:
\begin{align}\label{bcb}
|\overline{b}(t_0)| \gtrsim N_e \, \sqrt{2\epsilon} \, M_{\rm Pl}\, = \mathcal O(10^2)\,\sqrt{\epsilon}\, M_{\rm Pl}\,,  
\end{align}
with $N_e= \mathcal O(60-70)$ the number of e-foldings of the RVM inflation. The reader should also have noticed from \eqref{totalenerden}, that, as a result of the CS gravitational anomaly terms, the coefficient of the $H^2$ term in \eqref{totalenerden} is negative, in contrast to the standard form, current-era, RVM energy density \eqref{rvmener}, in which $\nu > 0$. As we have already mentioned, such negative contributions to the energy density are characteristic of higher-curvature gravities, of which CS gravity is an example (Gauss Bonnet-dilaton gravity~\cite{kanti} is another example, as already mentioned). 
 
The EoS of the string-inspired model during this inflationary phase has been explicitly computed in
\cite{ms} and found to coincide with the RVM EoS \eqref{rvmeos}, which justifies {\it a posteriori} calling this model {\it stringy RVM}.  Because of the RVM form of the vacuum energy density \eqref{totalenerden}, we observe that it is the fourth power $H^4$ of the Hubble parameter, which is dominant in the early Universe, that drives inflation in this model, which thus is of RVM form, not requiring external inflaton fields for its realization. To see this, we consider the RVM evolution~\cite{lima}, based essentially on conservation of the total energy-stress tensor, which is valid in our CS cosmology as well, as a result of a Bianchi identity for the Einstein tensor. Indeed, the graviton (Einstein) equations obtained from \eqref{sea3}, read~\cite{bms}:
\begin{align}\label{einsteq}
R^{\mu\nu} - \frac{1}{2}\, g^{\mu\nu}\, R = \sqrt{\frac{2}{3}}\, \frac{\alpha^
\prime \, \kappa}{12} \, C^{\mu\nu} + \kappa^2 \, T^{\mu\nu} \equiv \mathcal T^{\rm tot \, \mu\nu}\,,
\end{align}
where the Cotton tensor is not covariantly conserved~\cite{jackiw}, $C^{\mu\nu}_{\,\,\,\,\,\,\,;\mu} = -\frac{1}{8}\,( \partial^\nu b ) R_{\mu\nu\rho\sigma}\, \widetilde R^{\mu\nu\rho\sigma}$, which expresses an exchange of energy between the axion and the gravitational anomaly terms, leading to the aforementioned negative contributions of the CS anomaly to the stress-energy tensor $T^{\mu\nu}$ of the system. Nonetheless, as a result of the Bianchi identity for the Einstein tensor (that is, the vanishing of the left-hand side of \eqref{einsteq} upon contracting \eqref{einsteq} with the (torsion-free) gravitational covariant derivative $;\nu$), the total stress-energy tensor  $\mathcal T^{\rm tot \, \mu\nu}$ is conserved, 
\begin{align}\label{totconsstress}
{\mathcal T^{\rm tot \,\,\mu\nu}}_{\,;\nu}=0\,,
\end{align}
and this leads to the RVM evolution for the Hubble parameter~\cite{lima}:
\begin{align}
\label{HE}
{\dot H}+\frac{3}{2}(1+\omega_i)\, H^2
\left(1-\nu_{\rm infl}- 
\alpha_{\rm infl} \, \frac{H^2}{H_{I}^2}\right)=0,
\end{align}
where $\nu_{\rm infl} =  -\frac{1}{6}\, \epsilon < 0$ and  $\alpha_{\rm infl} = \frac{4.3}{3}\times 10^{10} \, \sqrt{\epsilon} \, \frac{|\overline b(0)|}{M_{\rm Pl}^3}\, H_I^2 > 0$ in the stringy RVM case. The quantity 
$\omega_i=\rho_{i}/p_{i}$ is the equation-of-state parameter for the pertinent ``matter'' fluid components $i$ that dominate in our early string-inspired universe in the era of interest, with $\rho_i$ ($p_i$) the appropriate energy (pressure) density, which correspond to appropriate components of $T^{\mu\nu}$ in \eqref{einsteq}. In the conventional RVM, the index $i$ runs successively over the radiation and then matter epochs, and corresponds to the pertinent excitations above the vacuum energy. In our strting-inspired cosmology, $i$ pertains to the fluid excitations that involve the KR axion and the fluctuations of the CS terms, that lead to the Cotton tensor. These are up and above the  
condensate vacuum energy contributions. A solution of the evolution equation \eqref{HE} is given by~\cite{lima}:
\begin{align}
\label{HS1}
 H(a)=\left(\frac{1-\nu_{\rm infl}}{\alpha_{\rm infl}}\right)^{1/2}\,\frac{H_{I}}{\sqrt{\mathcal D_0\,a^{3(1-\nu_{\rm infl})(1+\omega_i)}+1}}\,,
\end{align} 
where $a(t)$ is the Universe scale factor, and $\mathcal D_0 > 0$ is an integration constant, to be fixed by the appropriate boundary conditions at the interface between the 
onset of the RVM inflationary era and the preceding epoch, which in the model of \cite{ms} is dominated by the stiff KR axion matter. The consistency of the inflation occurring at early eras, for which $a(t) \ll 1$, is evident from \eqref{HS1}, from which one obtains an approximately de Sitter era, with almost constant $H$.  
In the absence of CS anomalies, which is the case of the pre-RVM inflationary phase in the scenario of \cite{ms}, 
the equation of state of the KR axion ``matter'' without a potential, which dominates that era, is that of  a ``stiff'' matter $w_b=+1$.\footnote{We remark~\cite{ms} that cosmologies with stiff-matter-dominance eras that characterise the very early universe after the Big Bang also appear, but in a different context, in \cite{zeld}.} However, after the GW condensation, which result in the appearance of non trivial CS contributions, 
 the interactions between the KR-axion and the CS-anomaly make the equation of state of the fluid to be that of a de Sitter spacetime, $w_i \simeq - 1$, as the explicit computations of  \cite{ms} for that era have shown. The condensate contributions are also characterised by a de Sitter equation of state, and so the entire system is characterised by an RVM behaviour \eqref{rvmeos}. We note~\cite{ms} that, in the absence of the condensate, the system would have been characterised by a phantom-type de Sitter energy~\cite{phantom}, $\rho_{\rm vac}=-p_{\rm vac} <0$. Hence, it is the {\it dominance} of the CS {\it condensate} that leads to positive RVM type vacuum energies $\rho^{\rm total}_{\rm vac}=-p^{\rm total}_{\rm vac} > 0$. 
 
 Thus, from \eqref{HS1}, we obtain
an approximately constant $H(a)|_{\rm RVM~infl} \simeq \left(\frac{1-\nu_{\rm infl}}{\alpha_{\rm infl}}\right)^{1/2}\,\frac{H_{I}}{\sqrt{\mathcal D_0\,+1}}$, consistent with an inflationary era. Identification of $H(a)|_{\rm RVM~infl}$ with $H_I$, fixes the constant $\mathcal D_0= \frac{1-\nu_{\rm infl}}{\alpha_{\rm infl}} -1 \simeq \frac{3}{4.3 \, \epsilon} -0.88 $, using 
\eqref{bcb}. From the Friedman equation then, using \eqref{totalenerden}, we obtain~\cite{ms} $\epsilon = \mathcal O(10^{-2})$, which yields $\mathcal D_0 \gg 1 > 0$, as expected. This proves the self consistency of our approach as far as the inflationary era is concerned, which arises from the RVM evolution. We note that, as evidenced from \eqref{HS1}, the RVM inflation is driven by the non-linear $H^4$ terms, dominant at early eras of the Universe, and it is {\it not} due to external inflaton fields.

\section{Primordial Black Holes during the stringy RVM inflation and post inflationary gravitational-wave profiles} \label{sec:pbh}

In this section we shall study the consequences of the presence of periodic modulations of the potential of axion fields that arise from compactification in string theory, and co-exist with the KR axions~\cite{svrcek,arvanitaki}.  We shall argue that such structures enhance the densities of pBH  which are produced from the vacuum during the RVM inflation~\cite{pbH}. Such an effect can in turn alter the profiles of GW during the radiation era that succeeds the RVM inflation, leading to in-principle observable patterns in the respective power spectra~\cite{stamou}. In addition, the enhanced pBH densities can imply the potential r\^ole of pBHs as a DM component~\cite{stamou,pbH}. 

Before commencing our analysis it is useful to place our work on a more general, string-related, perspective. The formation of gravitational CS anomaly condensates in \eqref{sea3} leads to a linear $b$-axion potential, formally similar to the one characrterising the so-called {\it axion-monodromy} string-inspired models of inflation~\cite{silver}. In such models, the stringy axion fields $a(x)$, deriving from brane compactifications in, e.g. type IIB string theory,, are characterised by a linear potential density, which drives inflation, 
\begin{align}\label{linearaxion}
V(a) = \Lambda_0^3 \, a(x)\, ,
\end{align}
where the dimensionful constant $\Lambda_0$, with mass dimension +1, is related to parameters in the underlying brane theory model. Specifically, if one considers a D5 brane wrapped on a two cycle $\Sigma_I^{(2)}$ of size $\ell \sqrt{\alpha^\prime}$ (in the notation of \cite{silver}), then one obtains a corresponding potential for the axion field $a(x)$:
\begin{align}\label{axmonodr}
V(a) = \frac{\tilde \varepsilon}{g_s\, (2\pi)^5\, {\alpha^\prime}^2}\, \sqrt{\ell^4 + a^2(x)}\,,
\end{align}
with $g_s$ denoting the string coupling, and the constant parameter $\tilde \varepsilon$ being associated with effects of warping. For large axion fields, compared to the mass scale $\ell$ one then obtains from \eqref{axmonodr} a {\it linear} axion potential, a phenomenon known as axion monodromy, which is common in string compactifications~\cite{silvermon}.

There are formal similarities, but also essential differences, between our stringy RVM scenario~\cite{bms,ms} and the axion monodromy case~\cite{silver}  \eqref{linearaxion} in the large $a(x)/\ell^2 \gg 1$ limit. The first difference of the string model from our stringy RVM scenario is that the linearity of 
the potential \eqref{axmonodr} is an approximate situation, which leads to linear axion potential in terms of the absolute value of the large axion field $|a(x)| >0$. Par contrast, the linear axion potential in the stringy RVM~\cite{bms,ms} is due to the {\it exact} coupling of the axion field to the gravitational anomaly CS term, with the linearity arising when condensates of  this anomaly form, {\it e.g.} as a result of primordial chiral GW, as we discussed above ({\it cf.} \eqref{condensateN2}). However, in contrast to the string axion-monodromy case, in which it is the axion that drives inflation, and thus the axion can play the r\^ole of an inflaton field, in the stringy RVM, the condensate is not quite a constant, but instead depends on the Hubble parameter $H$, which during inflation varies mildly with the cosmic time, and moreover leads to non-linear $H^4$ terms in the corresponding vacuum energy density \eqref{totalenerden}, which constitute the driving force for the RVM inflation, without the need for external inflatons~\cite{bms,lima}. Nonetheless, there are some features that the stringy RVM can share with the axion-monodromy string models, namely the existence of axion-potential modulations, arising from world-sheet non-perturbative effects, which we now proceed to discuss.

\begin{figure}[ht]
\center
\includegraphics[width=20pc]{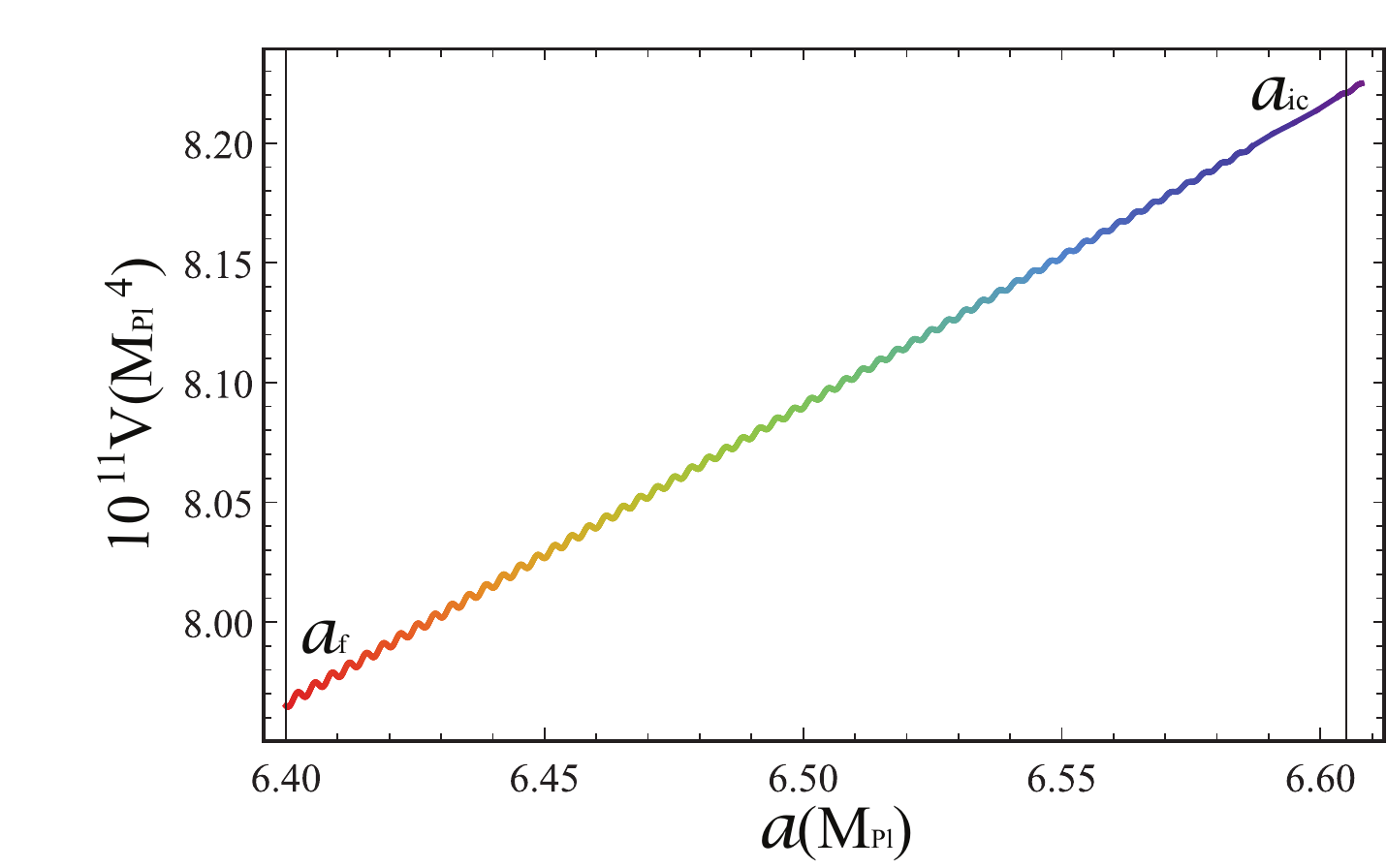}
\caption{The  potential \eqref{effpot} in the $a$-field direction for the set of parameters
of Table \ref{table1}. The quantity $a_{ic}$ gives the initial condition of the $a$ axion, at an onset time 
before its oscillations start to become appreciable, whilst $a_{f}$ corresponds to the value of the field when the inflation ends. Picture taken from \cite{stamou}.}
\label{fig:pot}
\end{figure}

Specifically, below we shall demonstrate under what conditions such features can characterise a variant of the stringy RVM, examined in \cite{stamou}. In the context of the string-effective action \eqref{sea3}, we consider the r\^ole of a second axion, arising, e.g., from some sort of compactification in the underlying microscopic string theory. As a concrete phenomenological example, we assume that these two axions, the string-model-independent KR one, $b(x)$, and the compactification axion $a(x)$, are the only axions which have dominant effects during the RVM inflationary phase, which is {\it driven} by the $b(x)$ axion. ~The second axion $a(x)$ is held responsible for {\it prolonging} the inflationary phase. It is this second axion which is assumed to have periodic modulation structures in its potential, as a consequence of some world-sheet non-perturbative effects. The 
two-axion potential in such a model assumes the form~\cite{stamou}:
\begin{equation}\label{effpot}
 V(a, \, b)={\Lambda_1}^4\left( 1+ f_a^{-1}\, \tilde \xi_1 \, a(x) \right)\, \cos({f_a}^{-1} a(x))+\frac{1}{f_{a}}\Big(f_b \, {\Lambda_0}^3 + \Lambda_2^4 \Big) \, a(x) + {\Lambda_0}^3\, b(x), 
 \end{equation}
 where $f_a$ is the axion-$a$-field coupling, the KR axion coupling $f_b$ is given by \eqref{fbdef}, 
and, in the study of \cite{stamou}, the parameters $\tilde \xi_1, f_a$, and the world-sheet-instanton-induced scales $\Lambda_1, \Lambda_2 $, are treated as phenomenological, with the constraint that:\footnote{For practical purposes, such a hierarchy of scales \eqref{constr} implies that the term $\Lambda_2^4$ can be ignored in front of $\Lambda^3_0 f_b$ in \eqref{effpot}.}
\begin{align}\label{constr}
 \Big(\frac{f_b}{f_a} + \frac{\Lambda^4_2}{f_a\, \Lambda^3_0} \Big)^{1/3}\, \Lambda_0  \, < \, \Lambda_1 \ll \Lambda_0 ~.
 \end{align}
 so that the RVM inflation is driven by the $b(x)$ field, thus maintaining the spirit of the stringy RVM~\cite{bms,ms}.
 As already mentioned, the $a(x)$ field is held responsible for only prolonging the RVM inflation.
 Both axions exhibit a linear term in their potential, due to their coupling with the CS gravitational anomaly via their respective axion-coupling parameters, which undergoes condensation. In the context of the stringy RVM, such CS  condensates are represented by the (approximately) constant quantity 
 \begin{align}\label{cond2}
 \Lambda_0^3 = \frac{1}{f_b} \langle R_{\mu\nu\rho\sigma}\, \widetilde R^{\mu\nu\rho\sigma}\rangle
 \end{align} 
 during the stringy RVM inflation.
 We stress that the periodic modulations of the axion-$a(x)$ potential in \eqref{effpot}, generated by the non-perturbative world-sheet effects, do not lead to mass terms for the axion $a(x)$. Moreover, we assume that the $b$ field is not characterised by such modulations. 
The potential \eqref{effpot} in the $a$-direction is plotted in fig.~\ref{fig:pot} for a concrete set of parameters, given in Table \ref{table1}~\cite{stamou}.  

\begin{figure}[ht]
\centering
\includegraphics[width=75mm,height=55mm]{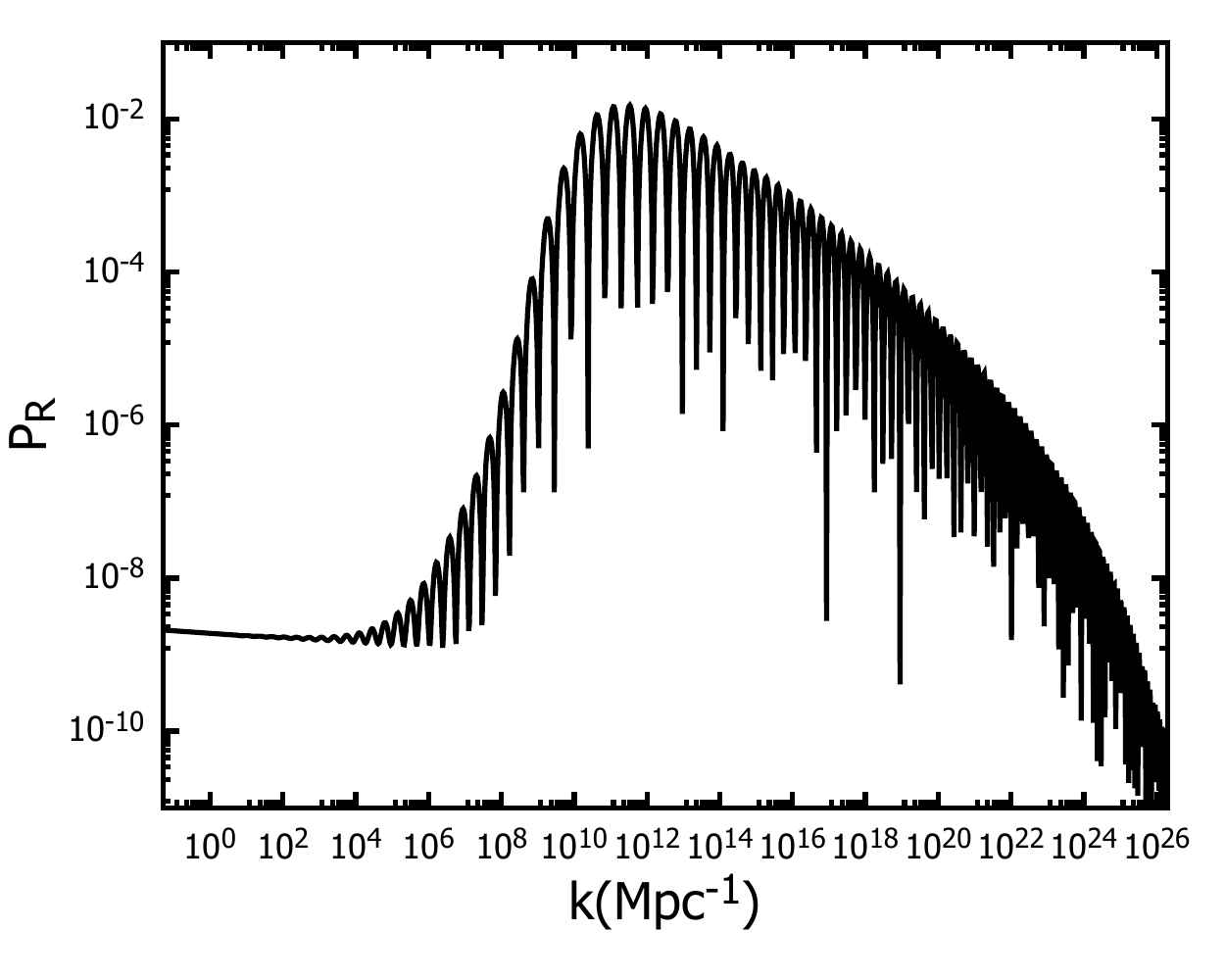}
\includegraphics[width=75mm,height= 63mm]{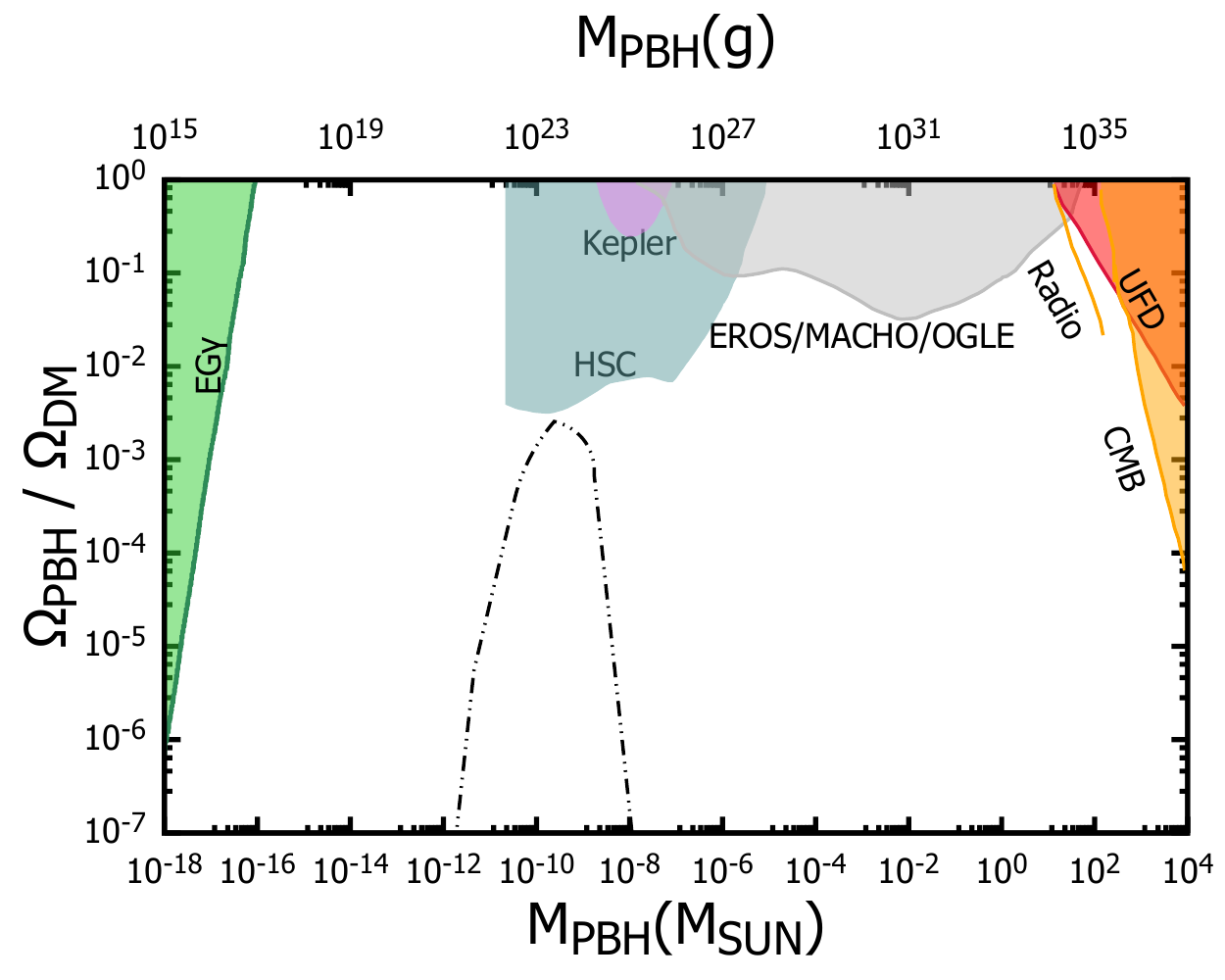} \\ \includegraphics[width=80mm]{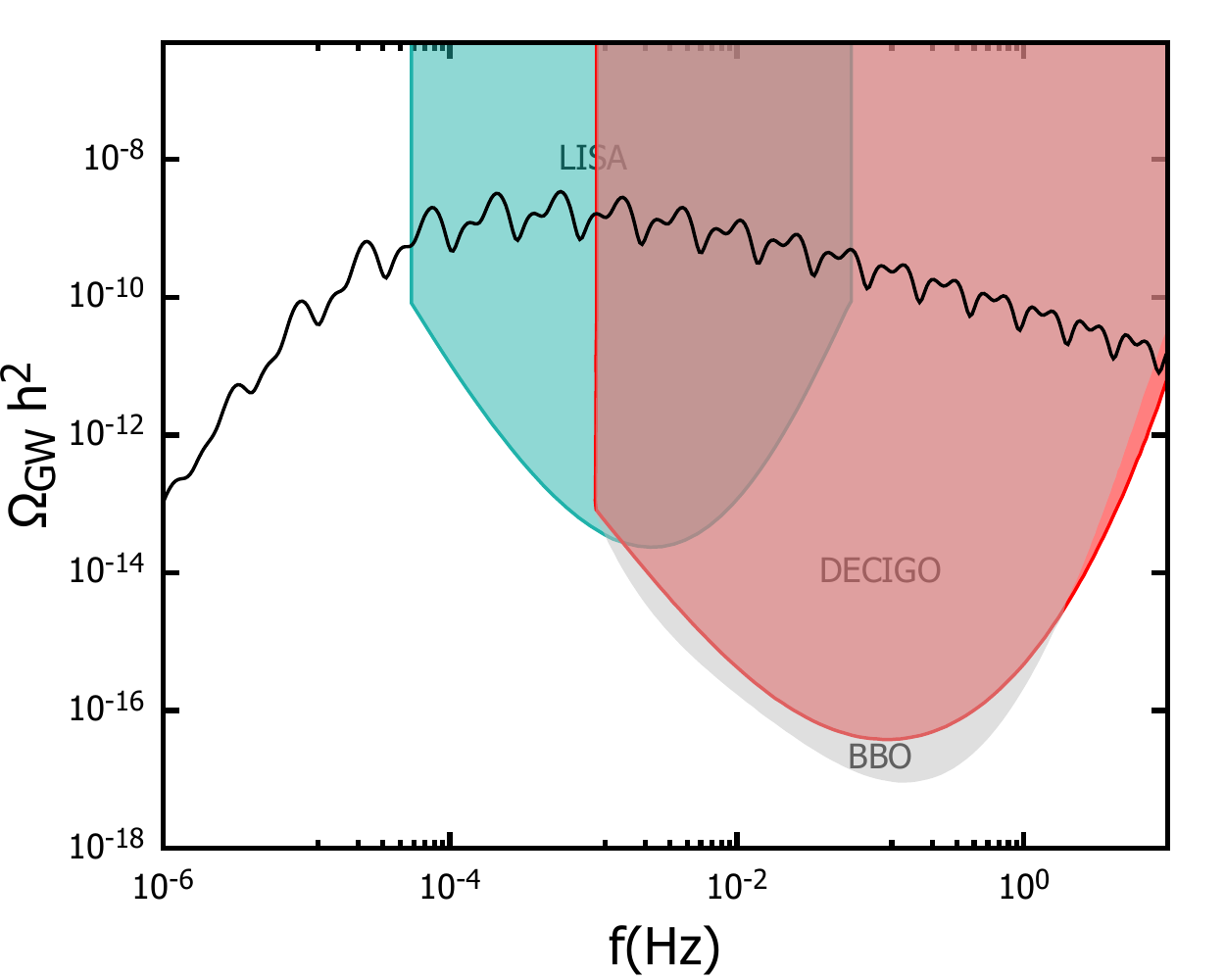}
\caption{ The results of the power spectrum, fractional abundance of PBHs and the energy density of induced GWs for the first set of parameters given in Table \ref{table1}.  We assume the following initial conditions for the axion fields (in units of $M_{\rm Pl}$): $a_{ic}=6.605$, and $b_{ic}= 11.100 $. The fractional abundance of pBH is  taken to be $f_ {PBH}=0.01$. Exclusion regions, including those from future interferometers (LISA) are shown. Similar features seem to characterised cases with higher values of $f_{PBH}$. Picture taken from \cite{stamou}.}
\label{fig:power}
\end{figure}

\begin{center}
\begin{table}[ht]
\begin{tabular}{||c| c c c c|c c c||} 
 \hline
SET & $g_1$ & $g_2$ & $\xi$ &$f(M_{Pl})$ &$\Lambda_0(M_{Pl})$ & $\Lambda_1(M_{Pl})$ &${\Lambda_3}(M_{Pl})$ \\ [0.5ex] 
 \hline\hline
 1 &$ 0.021$ & $0.904 $&$ -0.15$ & $2.5\times 10^{-4}$&$ 8.4 \times 10^{-4} $&$8.19\times 10^{-4}$ &$2.32\times 10^{-4}$\\
  \hline
\end{tabular}
\caption{An indicative example of parameters of the potential \eqref{effpot}~\cite{stamou}. We use the notation: $g_1 \equiv \frac{f_b}{f_a} + \frac{\Lambda^4_2}{f_a\, \Lambda^3_0}  \,, \,
\Lambda_1^4 \equiv g_2\, \Lambda_0^4\, , \,f \equiv f_a\,, \,  \xi \equiv  \frac{M_{\rm Pl}}{f_a}\, \tilde \xi_1\,.$ }
\label{table1}
\end{table}
\end{center}
In this scenario, the KR field $b$ dominates and drives the first stage of (RVM) inflation, 
when the oscillations of the axion $a$ are suppressed compared to its linear potential term.  As the cosmic time elapses, the value of $b$ decreases~\cite{stamou}, and eventually the oscillations of the  field $a$ become dominant, 
appearing effectively as many small  ``step''-like patterns in the potential. Such structures lead to an enhancement of the densities of pBH that are produced from the gravitational vacuum during inflation, and are imprinted in the primordial spectra~\cite{stamou}. This has consequences in the profiles of GWs in the radiation era, which succeeds the RVM inflation~\cite{lima,bms,ms}, in a similar spirit to what happens in the case of inflationary potentials with discontinuities~\cite{tetradis}. However, in contrast to those scenarios, the evolution here is smooth, and there are no discontinuities. The detailed analysis of \cite{stamou} has demonstrated that such enhanced production of pBH can lead to significant fraction of the DM in the Universe being due to pBHs~\cite{pbH}.
The situation is depicted in fig.~\ref{fig:power} for a concrete set of parameters and fractional abundance of pBH, $f_{PBH}=0.01$.

\begin{figure}[ht]
\centering
\includegraphics[width=75mm,height= 55mm]{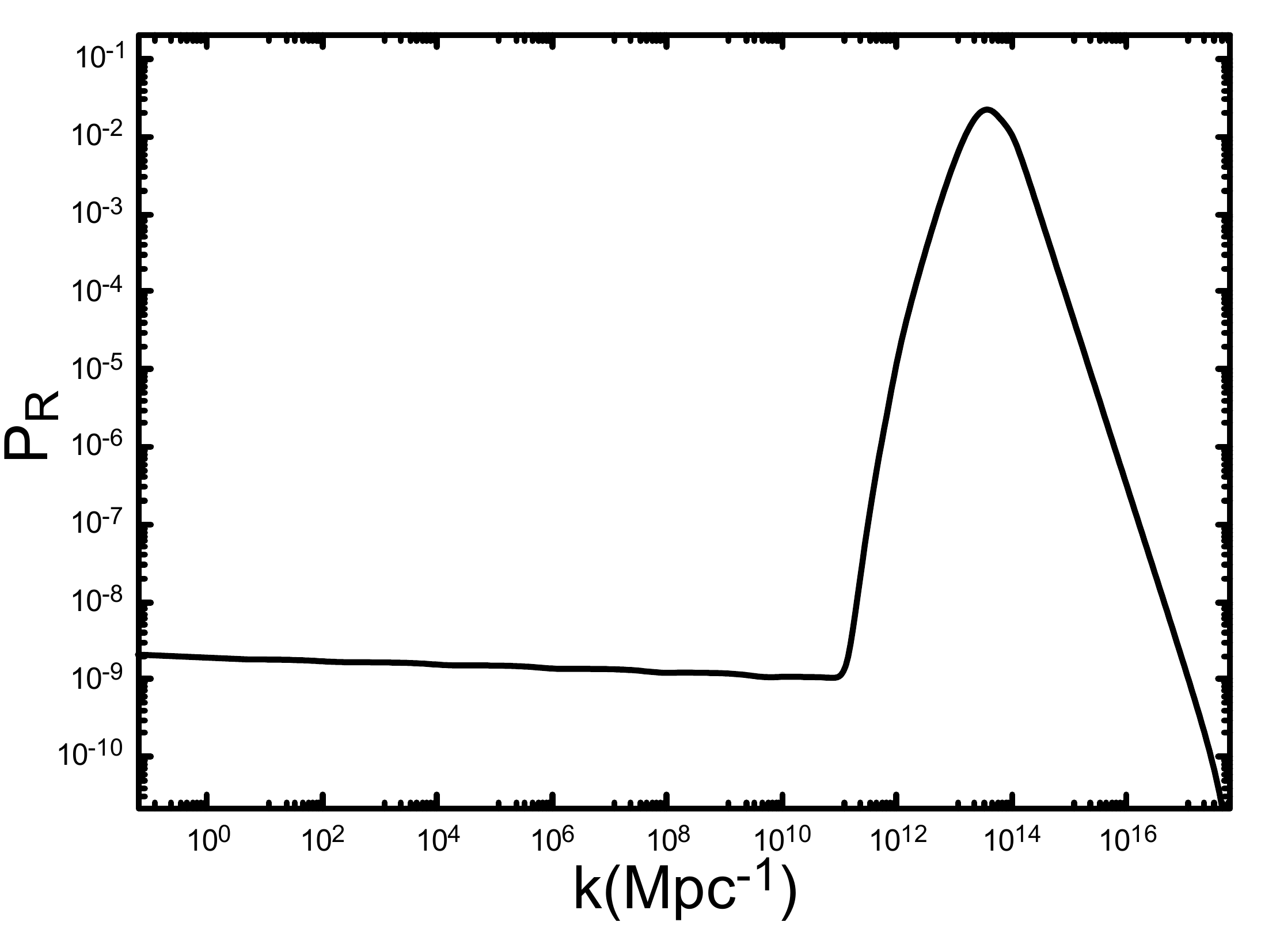}
\includegraphics[width=75mm,height= 63mm]{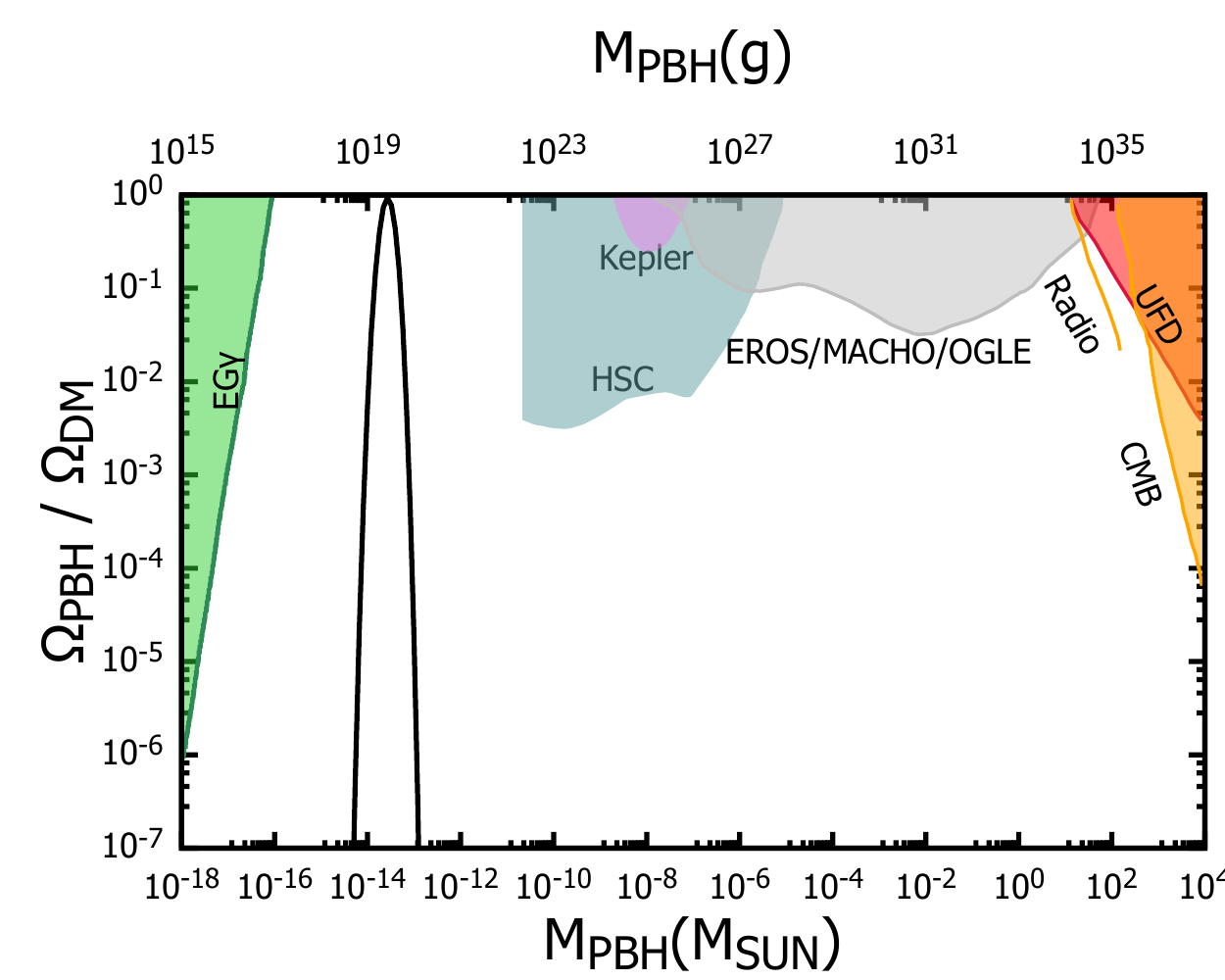}\\
\includegraphics[width=80mm]{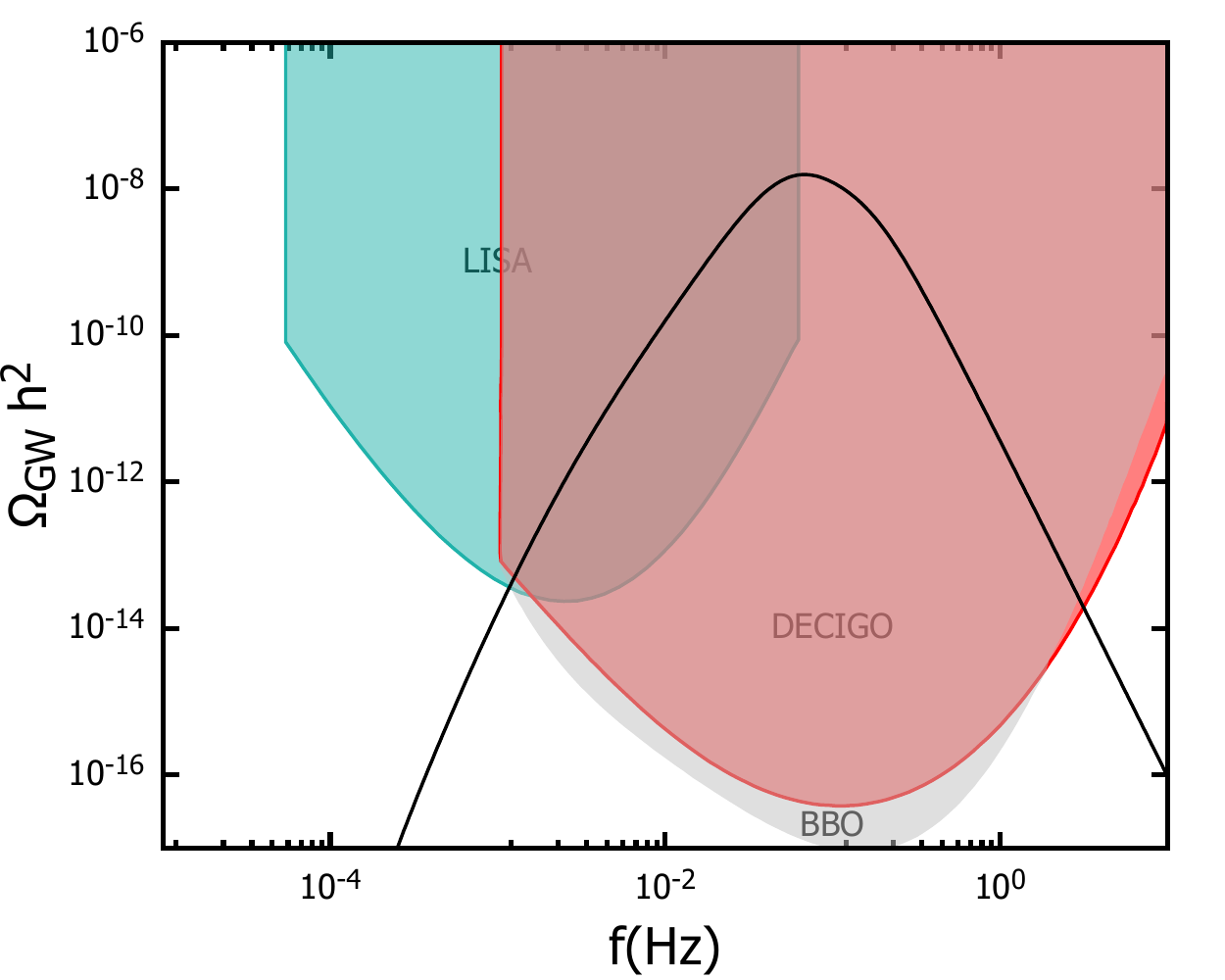}
\caption{\underline{Upper left panel}: The  power spectrum for the model of \cite{sasaki}, with the hierarchy of scales \eqref{constr3}.  \underline{Upper right panel}: The abundances of pBHs. \underline{Lower panel}: The energy density of GWs. The diagrams correspond to the following choice of parameters $\Lambda_0=8.4\times 10^{-4} M_{\rm Pl},  \quad  \Lambda_1=9.7 \times 10^{-3} M_{\rm Pl}\, , \,g_1=110\, , \, g_2=1.779\times
10^4\, ,\, \xi=-0.09\, , \, f=0.09 \,\, M_{\rm Pl}\,$, which obey the hierarchy \eqref{constr3}. 
 The axion fields  $a$, $b$ have the following initial conditions (in units of $M_{\rm Pl}$): $(a_{ic}\, , \, b_{ic})= (7.5622, \, 0.5220)$. Exclusion regions, including those from future interferometers (LISA) are shown. Picture taken from \cite{stamou}.}
\label{fig:featureless}
\end{figure}

These results should be compared with the ones characterising other string-inspired axion-monodromy models with two axions, considered in \cite{sasaki}, characterised by a different hierarchy of scales than \eqref{constr}:
 \begin{align}\label{constr3}
 \Lambda_0 \, \ll  \, \Big(\frac{f_b}{f_a} + \frac{\Lambda^4_2}{f_a\, \Lambda^3_0} \Big)^{1/3}\, \Lambda_0  \, < \, \Lambda_1 ~.
 \end{align}
In such models it is the axion $a$, whose potential contains non-perturbatively-induced periodic structures, that drives initially inflation, whilst the KR axion $b$ is responsible for prolonging it. As becomes clear from fig.~\ref{fig:featureless}, 
although such cases also lead to enhanced power spectra and enhanced production of pBH during inflation, nonetheless 
they are characterised by featureless (smooth) energy density profiles of GW after inflation, which
makes the models of \cite{sasaki} in principle distinguishable experimentally from the extended stringy-RVM model of \cite{stamou}, {\it e.g.} in future interferometers, such as LISA~\cite{LISA}. 

We finally remark that  the sets of parameters used in the above analysis and in \cite{stamou}
are such that there is no prolonged reheating period from the end of inflation till the onset of radiation.
In general, however, RVM models might be characterised by such prolonged reheating periods~\cite{lima}, which 
may lead to further enhancement of the power spectra and the densities of pBH. Moreover, as we have already mentioned, in realistic string models, many more axion fields, obtained from compactification~\cite{svrcek}, might come into play, thus affecting the fractions of pBH that could play the r\^ole of DM components, leading to a much richer phenomenology and early-Universe cosmology~\cite{arvanitaki,marsh}.

\section{Modern Eras, stringy RVM and cosmological tensions: speculations}\label{sec:now} 

A detailed scenario for the post inflationary eras of the stringy RVM has been described in \cite{bms,ms}, where we refer the interested readers. We only mention here that at the end of the RVM inflation, the decay of the metastable vacuum leads to the generation of chiral fermionic matter and radiation. The chiral matter generates its own gravitational CS, but also chiral global anomalies, associated with the gauge sector. The matter-generated gravitational anomalies  {\it cancel} the primordial ones due to the Green-Schwarz mechanism, but the chiral anomalies remain. It is possible then, as discussed in \cite{bms,ms}, that during the cosmic epoch corresponding to energy scales that correspond to dominance of Quantum Chromodynamics (QCD) effects, instanton effects of the colour $SU_{\rm c}(3)$ gauge group are responsible for generating periodic potentials for the KR (and also other) axions in generic string models, of the form:
\begin{align}\label{qcdpot}
V_b = \Lambda^4_{\rm QCD}\, \Big(1 - \rm cos\Big[\frac{b}{f_b}\Big]\Big)\,,
\end{align}
where $\Lambda_{\rm QCD}$ is the QCD scale, which may be assumed to be of order $\mathcal O(200)$~MeV, with the 
$f_b$ given in \eqref{fbdef}. We observe from \eqref{qcdpot} that a mass for the axion $b$ can be generated, which is of order $m_b = \frac{\Lambda^2_{\rm QCD}}{f_b} \sim 10^{-22} \Big(\frac{M_{\rm Pl}}{M_s}\Big)^2\, \rm GeV$.\footnote{In more precise estimates, $\Lambda_{\rm QCD}$ is replaced by the quantity $\sqrt{m_\pi\, f_\pi}$, where $m_\pi$ is the pion mass, and $f_\pi$ the pion decay contant. In such more precise treatments, the instanton-induced potential \eqref{qcdpot} is modified accordingly~\cite{kim}.} 	  
Depending on the value of the string mass scale $M_s$, which from collider experiments is known to be in the range $M_{\rm Pl} \gtrsim M_s \gtrsim \mathcal O(\rm 10)~TeV$, one may have axion masses  in a region from ultralight $10^{-11}$ eV to heavy, $10^{+11}$~eV, respectively.
Heavy axions are excluded from a plethora of astro-particle constraints~\cite{marsh}. For instance, non-derivative interactions of nucleons (or in general heavy fermions) $f$ with axions, may alter the proton to neutron ratio during BBN, 
 via processes ``axions + photons $\rightarrow f + \overline f$'', thus constraining the axion coupling to be $f_a < 10^9$~GeV~\cite{Conlon}, which, in the case of our QCD-like KR axion with coupling $f_b$  \eqref{fbdef},
 would imply  $m_b < 4 \times 10^{-2}$~eV, thus allowing for regions in which the KR axion mass assumes values in the range expected from QCD lattice considerations, for appropriate values of $M_s$. The requirement that such massive axions play a r\^ole as DM candidates imposes therefore phenomenological restrictions on the string mass scale $M_s$, which in our stringy RVM is kept arbitrary~\cite{mavlorentz}, equal to the ultraviolet cutoff of the effective low energy point-like gravitational field theory obtained from the underlying microscopic string theory. In this stringy RVM scenario, ultralight axions $a(x)$ of masses 
$m_a \lesssim 10^{-23}$, which could also play the r\^ole of DM, contributing to galactic growth~\cite{marsh}, provided they are the dominant DM species, cannot be the KR axion, but the compactification axions, see e.g.~\cite{nath} for generic scenarios of ultralight axions in string and supersymmetric theories. 
 
The interest in this section is the phenomenology of modern eras, during which, according to the data~\cite{Planck}, the Universe enters again a vacuum-energy dominated accelerating phase, approximating a de Sitter era. In the model of 
\cite{bms,ms} we still do not have a precise microscopic model that can explain this feature, although we offered plausible explanations within the RVM framework. Specifically, we have utilised chiral Abelian anomalies to argue that cosmic electromagnetic effects could condense in the current epoch and lead to a vacuum energy of almost de Sitter type. Moreover, due to the depletion of the current era of matter, gravitational anomalies, but much weaker than those during the inflationary era, could resurface~\cite{bms} and condense, thereby leading to a metastable, approximately de Sitter era now, of similar origin to the RVM inflationary one. Whatever the situation is, the RVM nature of our string-inspired cosmology is guaranteed, and in the current epoch one may parametrise the energy density of the vacuum as:
\begin{align}\label{modern}
\rho_{0\, \rm RVM} = \frac{3}{\kappa^2} \Big( c_0 + \nu_0 \, H^2 + \dots \Big)\,,
\end{align}
where the index $0$ denotes, as standard, present-day quantities, $c_0 >0$ is an approximate constant representing the current-epoch (metastable) cosmological constant, and $\nu_0 > 0$, as in the conventional RVM, which can be generated as a result of, say, cosmic electromagnetic effects~\cite{bms}. Higher (even) powers of $H$ are subdominant effects at late epochs, and hence we ignore them for the purposes of our discussion in this section.
The arguments of \cite{bms}, utilising cosmic magnetic fields of intensity $\mathcal B(t_0)$ that dominate the 
ground state today, lead to the following estimates for the cosmic rate of the KR axion background in the modern era:
\begin{align}\label{bdotmodern}
\dot b_{\rm today} \sim \sqrt{2\epsilon^\prime} \, H_0\, M_{\rm Pl}\,, \quad \sqrt{\epsilon^\prime} \simeq \frac{\sqrt{3}}{2}\, \frac{e^2}{4\pi^2}\, \frac{\mathcal B(t_0)}{k\, H_0\, M_{\rm Pl}}\,,
\end{align}
where $e$ is the positron charge, $H_0$ is the current value of the Hubble parameter, and $k$ is the momentum scale of the monochromatic cosmic electromagnetic fields~\cite{froh} that have been argued in \cite{bms} to provide sources for the currently-observed dark energy of the late RVM Universe.

Phenomenologically, in order to reproduce the correct amount of the observed DM composition in the Universe, we may expect~\cite{bms} $\epsilon^\prime $ to be of order $10^{-2}$, that is of the same order as the corresponding parameter $\epsilon$ that parametrises the cosmic rate of the KR axion background during the RVM inflationary era, {\it cf.} \eqref{axionbackgr}. A microscopic explanation of this `coincidence' feature is still lacking, given that estimates of the magnetic field intensity $\mathcal B (t_0)$ based on the underlying microscopic string theory are not available, and this quantity is considered as purely phenomenological in our approach.

Nonetheless, phenomenologically, a parametrisation of the RVM present-era energy density of the form \eqref{modern} leads to agreement~\cite{rvmpheno} with the current data upon using the phenomenological value $\frac{3}{\kappa^2}\, c_0 \sim 10^{-122} \, M_{\rm Pl}^4$. However, one also obtains appreciable small deviations from the $\Lambda$CDM, due to the $\nu_0 \, H^2$ terms, with $0 < \nu_0 = \mathcal O(10^{-3})$ as already mentioned, which fits the available data well. 
Most importantly, such a parametrisation can also provide~\cite{rvmtensions} an alleviation of the $H_0$ tension~\cite{tensions}, and, upon assuming a mild cosmic time $t$ dependence of the gravitational constant $\kappa \to \kappa(t)$ in \eqref{modern}, which leads to the so-called type II RVM model, one may be able to alleviate {\it simultaneously} the $H_0$ and $\sigma_8$ tensions. 

We now remark that, in the context of our stringy RVM model, as discussed in \cite{ms,mavrophil}, integrating out graviton quantum fluctuations  in the path integral around approximately (late-epoch) de Sitter cosmological space-times, one obtains logarithmic one-loop corrections to the energy density \eqref{modern}
of the form 
\begin{align}\label{qglncorr}
\delta \rho_{\rm RVM}^{\rm QG}  \propto d_1\, H^2\, {\rm ln}(H^2)\,, 
\end{align}
where $H(t)$ is the mildly depending on the cosmic time $t$ Hubble parameter in the approximately de Sitter current epoch. The coefficients $d_1$ are appropriate constants,
which are proportional to 
\begin{align}\label{cs}
 \tilde d_1 \propto \kappa^2 \mathcal E_0, \,\, {\rm or} \,\,\, \tilde d_1 \propto \kappa^2 \mathcal E_0 \, {\rm ln}(\kappa^4 |\mathcal E_0|)\,, 
\end{align}
with $\sqrt{|\mathcal E_0|}$ a bare (constant) vacuum energy density scale.  From \eqref{cs} it becomes clear that the sign of the coefficients $\tilde c_i, i=1,2$ depends on the signature of the bare cosmological constant term. In supergravity models $\mathcal E_0 < 0$, but in the absence of supersymmetry, as is the case of modern eras we are interested in, $\mathcal E_0$ could be positive. This will play an important r\^ole for the generic parametrization of a phenomenological analysis of the stringy RVM at late epochs~\cite{gms}. 

We mention at this stage that such logarithmic corrections to the vacuum energy density {\it also} appear in quantum field theories in Robertson-Walker space times, under appropriate renormalization group treatments~\cite{qftrvm}. 
Such matter-generated effects are expected to be present in the current era, since there is still appreciable matter content today ($\Omega_m \simeq 0.26$). It is therefore plausible that they have appreciable effects in 
alleviating the cosmological tensions. Naively, one would expect that such matter effects dominate over the quantum gravity effects. 

However, upon closer examination and comparison with \eqref{cs}, we see that this is highly dependent on the magnitude of $\mathcal E_0$. For concreteness let us consider the case of 
scalar real matter fields of mass $m$, non-minimally coupled to gravity with a parameter $\xi \in \mathbb R$~\cite{qftrvm},
with the conformal theory corresponding to $\xi=1/6$. In this case one obtains for the renormalised current era energy density in an expanding universe  spacetime with Hubble parameter $H(t)$:
\begin{align}\label{qftsc}
\rho_{\rm RVM}^{\rm vac~QFT}  (H) = \rho_{\rm RVM}^0 + \frac{3\, \nu_{\rm eff} (H) }{\kappa^2} \, \Big(H^2 - H_0^2 \Big), \quad 
\nu_{\rm eff} (H) \simeq \frac{1}{2\pi} \Big(\xi - \frac{1}{6}\Big) \, (\kappa^2 \, m^2)\,  \, {\rm ln}\Big(\frac{m^2}{H^2}\Big)  \,
\end{align}
The quantity $\rho_{\rm RVM}	^0 = 
\frac{3}{\kappa^2}(c_0 + \nu_0  \, H_0^2)$ is the standard current-era RVM energy density~\cite{rvmpheno}. 
The result \eqref{qftsc}, is based on the assumption~\cite{qftrvm} $|{\rm ln}(m^2/H^2)| \gg 1$. Qualitatively similar expressions characterise the fermion case~\cite{qftrvm}. However, the reader should notice that the corrections \eqref{qftsc} differ from the quantum graviton logarithmic corrections to the energy density, in that the logarithms 
appear in the combination 
\begin{align}\label{qftlncorr}
\delta \rho_{\rm RVM}^{\rm QFT} \propto (H^2-H_0^2)\,(\kappa^2 \, m^2){\rm ln}\Big(\frac{m^2}{H^2}\Big)\,, 
\end{align}
which are small in the modern era for which $H^2 \to H_0^2$. Unlike their quantum-gravity counterparts, such corrections cannot be cast in the form $R\,{\rm ln}(\kappa^2 R)$. However, comparing \eqref{qftlncorr} with \eqref{qglncorr}, \eqref{cs}, we observe that matter effects on the running vacuum energy density at modern eras could, depending on the magnitude of the bare cosmological constant $\mathcal E_0$, be suppressed compared to their quantum-graviton counterparts, given the smallness of the factor $(\kappa^2 \, m^2)\, (H^2 - H_0^2) \ll (\kappa^2 \, m^2)\, H^2 $ at modern epochs for which $H \to H_0$. Thus, even in modern eras, a non negligible possibility exists that quantum gravity corrections to the energy density of the stringy RVM dominate over quantum-matter-induced effects.

Making the latter assumption, we remark that such corrections can be best parametrised by an effective gravitational model with effective action~\cite{mavrophil}
\begin{equation}\label{eq:action1}
S=-\int d^4x\,\sqrt{-g}\,\left[c_0+ \frac{1}{\kappa^2}\, R\left(c^\prime_1+c_2\ln\left(\kappa^2 R\right)\right)\right]+S_m\,,
\end{equation}
where $S_m$ denotes the matter action and $c_1^\prime$, $c_2$ real constants, with $c_2$ a parametrisation of the logarithmic 
corrections due to (dominant) quantum gravity corrections.  
Since Physics should not depend on the units of the argument of the logarithm, one may consider instead 
$c_2\,{\rm ln}(\kappa^2 \, R) = c_2\,{\rm ln}\Big(\frac{R}{R_0}\Big) +  c_2\,{\rm ln}(\kappa^2 \, R_0)$, where $R_0=12\, H_0$ is the de Sitter curvature scalar today, and then absorb any dependence on the units of the argument of the logarithm in an appropriate redefinition of the constant  coefficient of the scalar curvature $R$ term in the action \eqref{eq:action1}, $c_1 \equiv c^\prime_1  +  c_2\,{\rm ln}(\kappa^2 \, R_0)$. 
Then, the parameter $c_1 = 1 + \delta c_1$, with $\delta c_1$ describing quantum corrections to the Newton's constant. Unitarity of the gravity sector requires $c_1 > 0$ in our conventions. This is the parametrisation we shall follow from now on, which leads to the effective gravitational action:
\begin{equation}\label{eq:action}
S=-\int d^4x\,\sqrt{-g}\,\left[c_0+ \frac{1}{\kappa^2}\, R\left(c_1+c_2\ln\left(\frac{R}{R_0}\right)\right)\right]+S_m\,.
\end{equation}
\begin{figure}[h]
\center
\includegraphics[width=30pc]{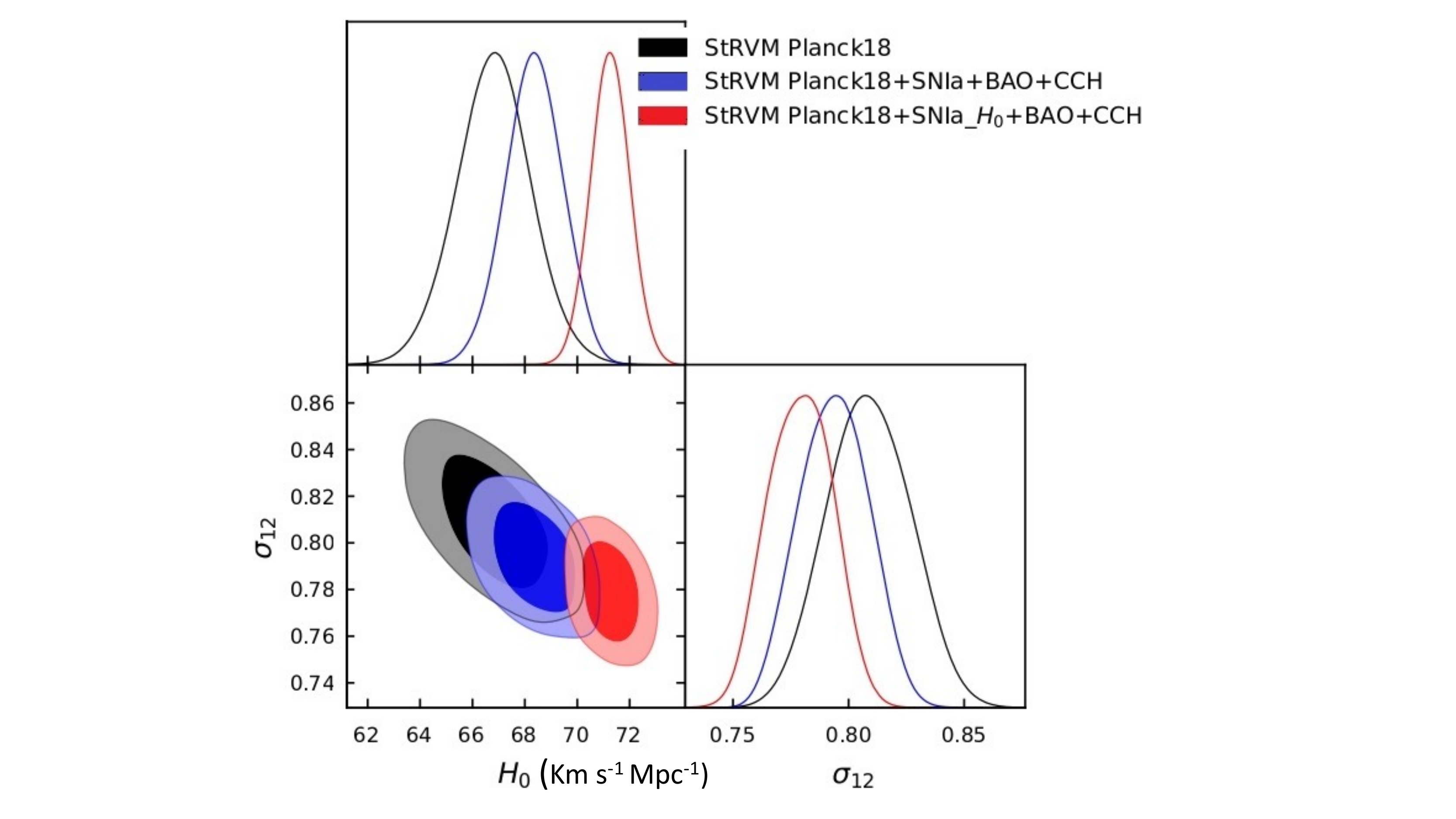} \hfill \includegraphics[width=20pc]{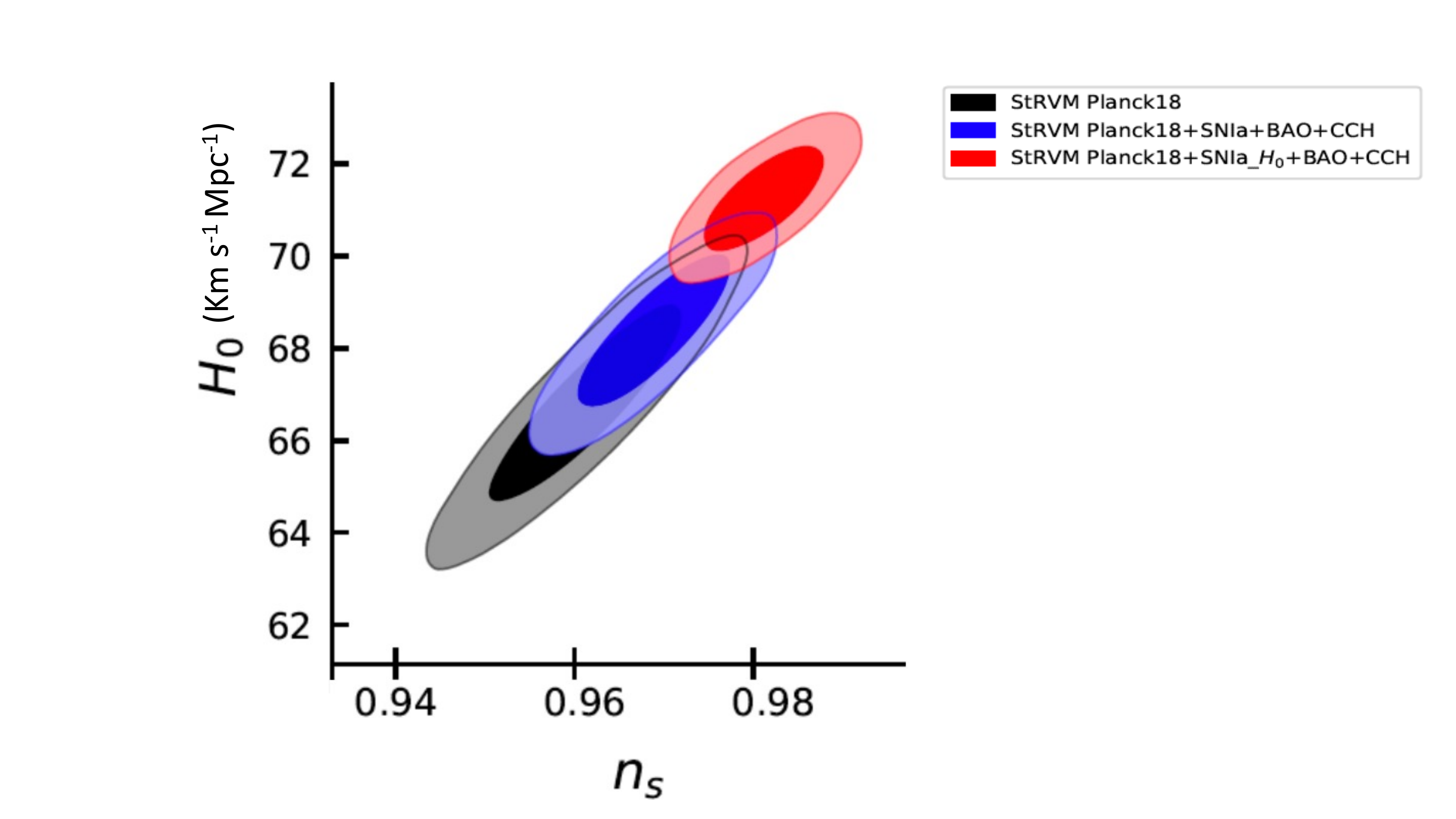} 
\caption{\underline{{\it Upper Panel}}: The alleviation of the $H_0$ tension and $\sigma_{12}$-growth tensions in the stringy RVM (StRVM) model of cosmology upon inclusion of graviton quantum fluctuations at one-loop order in modern eras, whose incorporation makes the corresponding effective gravitational action taking the form \eqref{eq:action}. \underline{{\it Lower Panel}}: The Hubble tension $H_0$ vs the spectral scalar index of primordial fluctuations, $n_s$, computed using standard parametrisation for inflation (in this latter respect we mention that we do not use the details of the stringy RVM inflation, for brevity, as they will not affect these conclusions). The various colours correspond to different combinations of cosmological data, as indicated on the top right panel of the figure panels. Figures taken from \cite{gms}.}
\label{fig:strvmH0}
\end{figure}
A full phenomenological analysis of this class of models, including fits to all available cosmological data at present (supernovae, CMB, Baryon Acoustic Oscillations and lensing data) is p[resented in ref.~\cite{gms}. 
On assuming that the supergravity contributions, corresponding to a primordial supersymmetry breaking scale $\sqrt{|f|}$, which contributes to the bare cosmological constant a negative value $-f^2 < 0$ (as required by supersymmetry~\cite{houston}), are the dominant ones of all the quantum-graviton-induced logarithmic corrections to the energy density from primordial to the current era,
so that the bare cosmological constant scale $\mathcal E_0$ appearing in \eqref{cs} coincides with $-\sqrt{f}< 0$, 
we have~\cite{ms,mavrophil,gms}:
\begin{align}\label{c12}
c_1 - c_2\, {\rm ln}(\kappa^2\, H_0^2) = 
\frac{1}{2\kappa^2} \Big[1 + \frac{1}{2}\, \kappa^4 \, f^2 \,\Big(0.083 - 0.049 \, {\rm ln}(3\kappa^4\, f^2)\Big)\Big]\, , \,\, c_2 = - 0.0045\, \kappa^2 \, f^2 < 0\,.
\end{align}
We then constrain the constants $c_1,c_2$ from data~\cite{gms}, in order to alleviate the 
$H_0$ and $\sigma_{12}$-growth tensions,\footnote{The reasons why for this stringy RVM model the data constrain better the parameter $\sigma_{12}$ rather than $f \sigma_8$ are explained in \cite{gms}.}  
with the situation being summarised  
in figure \ref{fig:strvmH0}, for the values of the dimensionless parameters  $|\frac{c_2}{c_1 + c_2}| = 
9 \times 10^{-3} \, \kappa^4 \, f^2 = \mathcal O(10^{-7})$ and $2 (c_1 + c_2) = 0.924 \pm 0.017$.
This yields the following estimate on the magnitude of $\sqrt{|f|}$, $\sqrt{|f|} \kappa \sim 10^{-5/4} < 1$, implying a subplanckian scale for primordial dynamical supergravity breaking, consistent with the transplanckian conjecture.
As follows from \eqref{c12}, such values are also consistent with the perturbative modifications of $c_1$ from the (3+1)-dimensional gravitational constant $1/(2\kappa^2)$, despite the fact that $\kappa^2 H_0^2 = \mathcal O(10^{-122}) \ll 1$. With such scales one can also see~\cite{mavrophil,ms} that the quantum graviton corrections during the RVM inflationary eras are subleading as compared to the $H^4$ terms in the vacuum energy density \eqref{totalenerden} induced by GW condensates, and thus they do not affect our mechanism for inflation discussed in section \ref{sec:rvminfl}. 

\section{Conclusions and Outlook}\label{sec:concl} 

In this talk, we have reviewed a string-inspired cosmological model with torsion and gravitational anomalies, 
which results in a Running Vacuum Model (RVM) cosmology. For this reason we term this model Stringy RVM. The model presents some interesting features. 

First, condensation of primordial gravitational waves, of chiral (left-right asymmetric) type, can lead to RVM inflation without the need for external inflaton fields, but being triggered by the $H^4$ terms in the vacuum energy, which owe their existence 
to the so-induced gravitational anomaly condensates, and dominate during the early eras. 

Second, the torsion degrees of freedom, which from a string theory point of view are associated with the field strength of the Kalb Ramond (KR) antisymmetric tensor field, give rise to the so-called string-model independent, or KR, axion. Other axions, arising from compactification to four dimensions, coexist with the KR axion. In such multiaxion models, and in particular in the concrete example of two axion models we examined above, in which only one of the compactification axions is dominant, it is possible that, within the framework of the stringy RVM, in which it is the KR axion that drives inflation, while the compactification axions may prolong its duration, there are periodic modulations of the potential of the compacitification axion induced by non-perturbative stringy effects. The presence of such periodic modulations might lead to an enhanced production of primordial black holes (pBH) during the RVM inflation, which leave their imprint in the profiles of the gravitational waves during the radiation epoch. Such imprints may be potentially detectable in future interferometers. 

The enhanced densities of pBH may also lead to their r\^ole as potential components of dark matter in significant fractions. On the other hand, the KR axions, might acquire non trivial masses during the post inflationary epochs, e.g. due to instanton-generated potentials in the epoch of dominance of Quantum Chromodynamics effects, thus playing the r\^ole of  another dark matter component. 

Finally, in modern eras of the cosmic evolution, the model also assumes an RVM behaviour, with a metastable vacuum energy, which has almost a de Sitter behaviour, and might be due to condensates of cosmic electromagnetic effects, or 
gravitational wave anomalies that resurface as a result of the depletion of matter. We have argued that quantum-gravity corrections due to 
quantum-graviton (tensor) fluctuations might lead to corrections in the vacuum energy, which, in addition to the $H^2$ terms, also involve $H^2\, {\rm ln}(H^2)$, and can dominate matter effects under some circumstances. The presence of such corrections might lead to an alleviation of the current-era cosmological tensions associated with $H_0$ and the $\sigma_{12}$ growth parameter, while the model is in agreement with the plethora of the other cosmological data, as detailed phenomenological analyses show~\cite{gms}.

\section*{Acknowledgements} I would like to thank the organisers of the Corfu 2022 workshop on {\it the Standard Model and beyond} for the invitation to speak in this interesting event. This talk is dedicated to the memory of two great colleagues, Prof. G.G. Ross and Prof. C. Kounnas, who have passed away unexpectedly. 
The work of NEM is supported in part by the UK Science and Technology Facilities research Council (STFC) under the research grant ST/T000759/1. N.E.M.  also acknowledges participation in the COST Association Action CA18108 ``{\it Quantum Gravity Phenomenology in the Multimessenger Approach (QG-MM)}''.

\end{document}